\documentclass[twocolumn]{aastex631}
\usepackage{natbib}

\newcommand{\CIV}{\ion{C}{4}}
\newcommand{\CII}{\ion{C}{2}}
\newcommand{\SiIV}{\ion{Si}{4}}
\newcommand{\SIV}{\ion{S}{4}}
\newcommand{\SiII}{\ion{Si}{2}}
\newcommand{\FeII}{\ion{Fe}{2}}
\newcommand{\PV}{\ion{P}{5}}
\newcommand{\CIII}{C~{\sc iii}*}
\newcommand{\NV}{\ion{N}{5}}
\newcommand{\Lyalpha}{Ly~{\sc $\alpha$}}
\newcommand{\msun}{\(\textup{M}_\odot\)}

\newcommand{\kms}{$\mathrm{km~s^{-1}}$}
\newcommand{\mbh}{$M_{\rm BH}$}

\newcommand{\simbal}{\textit{SimBAL}}
\newcommand{\swift}{\textit{Swift}}

\newcommand{\logU}{$\log(U)$}
\newcommand{\loga}{$\log(a)$}
\newcommand{\logNH}{$\log(N_{\rm H})$}
\newcommand{\logNHU}{$\log(N_{\rm H})-\log(U)$}
\newcommand{\cmsq}{cm$^{-2}$}

\submitjournal{ApJ}

\received{2022 October 07 }
\accepted{2023 June 12}

\usepackage{longtable}
\usepackage{graphics}
\usepackage{epsf}
\usepackage{graphicx}	
\usepackage{amsmath}	
\usepackage{amssymb}	
\usepackage{hyperref}

\shorttitle{Variable BALQ WPVS 007}
\shortauthors{K. Green}

\begin{document}

\title{Investigating the {Origin} of the Absorption-Line Variability in Narrow-Line Seyfert 1 Galaxy WPVS 007}
\author[0000-0003-2539-6884]{Kaylie S. Green}\footnote{Dr. Kaylie Green died tragically before the publication process for this paper could be completed.  The second author is the corresponding author for this publication.}\affiliation{Department of Physics and Astronomy, The University of Western Ontario, London, ON, N6A 3K7, Canada}
\author[0000-0001-6217-8101]{Sarah C. Gallagher}\affiliation{Department of Physics and Astronomy, The University of Western Ontario, London, ON, N6A 3K7, Canada} \affiliation{Institute for Earth and Space Exploration, The University of Western Ontario, London, ON N6A 3K7, Canada} \affiliation{The Rotman Institute of Philosophy, The University of Western Ontario, London, ON N6A 3K7, Canada}
\author[0000-0002-3809-0051]{Karen M. Leighly}\affiliation{Homer L. Dodge Department of Physics and Astronomy, The University of Oklahoma, 440 W. Brooks St., Norman, OK 73019, USA}

\author[0000-0002-3173-1098]{Hyunseop Choi}\affiliation{Homer L. Dodge Department of Physics and Astronomy, The University of Oklahoma, 440 W. Brooks St., Norman, OK 73019, USA}

\author[0000-0002-9961-3661]{Dirk Grupe}\affiliation{Department of Physics, Geology, and Engineering Technology, Northern Kentucky University, 1 Nunn Dr., Highland Heights, KY 41099, USA}
\author[0000-0002-0431-1645]{Donald M. Terndrup}\affiliation{Department of Astronomy, The Ohio State University, 140 W. 18th Ave., Columbus, OH 43210, USA}
\author[0000-0002-1061-1804]{Gordon T. Richards}\affiliation{Department of Physics, Drexel University, 32 S. 32nd St., Philadelphia, PA 19104}
\author[0000-0002-9214-4428]{S. Komossa}\affiliation{Max-Planck-Institut f{\"u}r Radioastronomie, Auf dem H{\"u}gel 69, D-53121 Bonn, Germany}

\begin{abstract}
Broad Absorption Line Quasars (BALQs) are actively accreting supermassive black holes that have strong outflows characterized by broad absorption lines in their rest-UV spectra. Variability in these absorption lines occurs over months to years depending on the source. WPVS 007, a low-redshift, low-luminosity Narrow-line Seyfert 1 (NLS1) shows strong variability over shorter timescales, providing a unique opportunity to study the driving mechanism behind this variability that may mimic longer scale variability in much more massive quasars. We present the first variability study using {the} spectral synthesis code \simbal, which provides velocity-resolved changes in physical conditions of the gas using constraints from multiple absorption lines. Overall, we find WPVS 007 to have a highly ionized outflow with a large mass-loss rate and kinetic luminosity. We determine the primary cause of the absorption-line variability in WPVS 007 to be a change in covering fraction of the continuum by the outflow. This study is the first \simbal\ analysis where multiple epochs of observation were fit simultaneously, demonstrating the ability of \simbal\ to use the time-domain as an additional constraint in spectral models.

\end{abstract}


\section{Introduction} \label{sec:intro}

Broad Absorption Line Quasars (BALQs) are the subset of quasars that have wide, blue-shifted absorption lines in their rest-UV spectra indicative of energetic outflows that can reach velocities of $\sim$0.1c \citep[e.g.,][]{1991Weymann}. BALQs make up roughly 20\% of the optically selected quasar population \citep{2003Hewett} and may impact the evolution of the host galaxy if their kinetic energies exceed 0.5--5\% of their rest-mass energy \citep[e.g.,][]{2010Hopkins}. Furthermore, BALQs may act to limit the growth of the central supermassive black hole by slowing accretion \citep[e.g.,][]{2005DiMatteo}.

Little is still known about how these outflows start and evolve, and the geometry of the outflow is poorly constrained as these sources are not spatially resolved. Studying the changes in outflow spectral signatures over time offers one method to map the outflow, for example by providing constraints on properties such as location of the outflow \citep[e.g.,][]{2017McGraw} and density of the gas \citep[e.g.,][]{1997Hamann} by making inferences about the cause of the variability. With the advent of new large quasar surveys there are multi-epoch observations available for many BALQs. These multi-epoch observations have shown (in some cases dramatic) variability of absorption-line structure. As different outflow velocities (with small velocities corresponding to small blueshifts from the central emission-line wavelength) could very well represent distinct locations within the wind, the rate and nature of variations can inform a picture of the geometry of the quasar wind and its driving mechanism, both of which are still poorly understood.

Broad quasar absorption lines frequently exhibit significant variation over rest-frame timescales from a few months \citep[e.g.,][]{Capellupo2013} to a few years \citep[e.g.,][]{Gibson2008}. Some absorption lines have even completely disappeared over observations spanning a few years \citep[e.g.,][]{2012Filiz}; the causes of such substantial changes in the spectra are still uncertain. Generally, the cause of variability has been attributed to either a change in the ionization state of the gas from a change in the continuum \citep{1993Barlow,2014FilizAk,2015Wang,Grier2015,2015Wildy,2018Rogerson,2019Hemler} or variation due to an absorber crossing transverse to our line of sight \citep{Gibson2010,2008Hamann,2012Filiz,2011Hall,Capellupo2013,2015McGraw,2017McGraw,2012Vivek,2016Vivek}.  Some studies are unable to distinguish between the two scenarios or have found support for both scenarios \citep{1987Foltz,Lundgren2007,Capellupo2012,2012Vivekb,2012Filiz,2014Vivek,2018Vivek}. To date, quantitative studies of variability have primarily relied on tracking changes in absorption-line equivalent widths {(EWs)} to determine the cause of the variability \citep[e.g.,][]{2012Filiz,Capellupo2013}, and the distribution of $\rm\Delta{EW}$ as a function of time can be used to obtain crude estimates of the physical conditions of the gas. 

These large-scale empirical studies have shown that higher-velocity components of BALs are more likely to vary than the low-velocity end of the absorption line \citep[e.g.,][]{Lundgren2007,Capellupo2011,2012Filiz}, and BALs with lower equivalent widths are more likely to vary \citep[e.g.,][]{1993Barlow,Lundgren2007}. Furthermore, BALs do not typically vary uniformly as a function of velocity \citep{Capellupo2012} and some lines (such as \SiIV\ $\lambda\lambda1393.8,1402.8$) are more likely to vary than others (e.g., \CIV\ $\lambda\lambda1548.2,1550.8$; \citealt{Capellupo2013}). 

With multiple epochs of observation, the nature of the absorption-line variability of a source can be used to estimate a lower limit on the density of the gas (and consequently an upper limit on the radius) under the assumption that the cause of the change is due to a change in the ionization state of the gas \citep[e.g.,][]{1993Barlow,FilizAk2013}. Alternately, if the cause of the change is due to an absorber moving across the observer's line-of-sight to the illuminating continuum, the radius of the outflow may be estimated from the crossing speed of the gas \citep[e.g.,][]{Capellupo2011}. Overall, the observations of variable absorption lines build up a dynamic picture of a wind that could either be reacting to changes in the ionizing spectral energy distribution resulting in changes in ionization parameter of the gas, or a cloud-crossing (eclipse) scenario where we see the depth of absorption lines change based on the gas column density and/or the fraction of the emitting source that is covered by the BAL gas. 

Currently missing from the ensemble of variability studies in the literature is a detailed study of the changing physical conditions of BAL features from multiple ions considered together with simultaneous UV photometry to provide a critical link between observed changes in the UV spectrum of a source and a change in the absorbing gas properties. Such a study has been difficult to perform for many BALQs because this would typically require broad wavelength coverage and BALs that are not blended in order to use template-fitting methods to extract the physical parameters. However, with the development of the novel spectral synthesis code \simbal\ \citep{Leighly2018}, we are able to use a forward modelling and spectral synthesis approach to spectral fitting, which allows the study of physical parameters for multiple blended lines simultaneously. The functionality of \simbal\ also allows physical parameters to be fit as a function of velocity, so that variation in the BAL can be studied as a function of ion species, epoch, and velocity.

An excellent choice for a case study as described above is WPVS 007, a Narrow-line Seyfert 1 (NLS1) galaxy at redshift 0.02861 (NASA/IPAC Extragalactic Database redshift value) with a small black hole mass ($\mathrm{4.1\times10^6\;M_{\odot}}$; \citealt{2009Leighly}) {and low luminosity ($\mathrm{5.20\times10^{43}\;erg\;s^{-1}}$; \citealt{Leighly2015}) that is behaving like a BALQ with a high-velocity outflow. Low-luminosity AGN with smaller black hole masses and size scales will show variability on smaller timescales than the much larger BALQs making a rare low-redshift source like \mbox{WPVS 007} an excellent laboratory to study variability in broad absorption line quasars. \mbox{WPVS 007} has a high velocity for its luminosity, making it an outlier for the velocity-luminosity relationship found by \citet{2002Laor} for low-redshift, soft X-ray weak quasars \citep[see Figure 13 and related discussion in][]{2009Leighly}.}

When WPVS 007 was first detected in the ROSAT All Sky Survey \citep{1999Voges}, it showed the softest x-ray spectrum ever observed in an AGN \citep{1995Grupe}.  Subsequent observations (including a \swift\ monitoring campaign from 2005 and continuing today) have also found the object to be X-ray-weak most of the time \citep{2007Grupe,2008Grupe,2017Komossa}. It is not visible in X-rays in a single \swift\ observation and requires merging many observations in order to obtain a detection. A 1996 HST UV spectrum showed only a mini-BAL (typical mini-BALs have FWHM greater than 200--300 \kms\ but less than 2000 \kms; \citealt{2004Hamann}); subsequently, a 2003 FUSE UV spectrum revealed the development of broad absorption lines ($\mathrm{V_{max}}\sim-6000$ \kms\ and FWHM $\sim3400$ \kms; \citealt{2009Leighly}) in addition to the low-velocity mini-BALs ($\mathrm{V_{max}}\sim-900$ \kms\ and FWHM $\sim550$ \kms; \citealt{2009Leighly}). Further UV spectroscopy with HST in 2013 indicated large variability in the broad absorption lines over short periods \citep[e.g.,][]{2013Grupe,Leighly2015}. \swift\ monitoring shows that WPVS 007 became fainter in UV (across \swift\ bands including U, V, B, UVW1, UVW2, and UVM2) from 2010 until 2015, reaching its faintest magnitude in February and March in 2015 before brightening again in 2017. These changes were determined to be an occultation event by \citet{Leighly2015} when the analysis of the photometry showed that the occultation dynamical timescale, the BAL variability timescale, and the density of the BAL gas taken together are consistent with the reddening material and the broad-absorption line gas originating in the torus. 

{More recently, \citet{Li2019} presented a detailed comparison of WISE mid-infrared photometry and the UV photometry from \swift. They attribute the variations in both data sets to changes in the accretion-disk luminosity and note that the brightness changes seen by WISE lag behind those seen in \swift\ by roughly 600 days as expected from the physical size of the UV and mid-infrared emitting regions at their respective wavelengths. They further propose that the strengthening and emergence of high-velocity absorption from the \SiIV\ and \CIV\ BALs are caused by an increase in the brightness of the ionizing continuum.}

{Our objective in this paper is to use the body of evidence in multiple epochs of UV spectroscopy to re-evaluate the cause of the BAL variability, taking into account velocity-resolved information.}  In \S\ref{sec:obs}, we present the five HST COS observations used in our analysis. The process used for fitting the continuum, emission-line and absorption-line features using Sherpa \citep{Sherpa2001} and \simbal\ is described in \S\ref{sec:methods}. The results from the fitting and the primary physical driver of the observed variability in the absorption lines of WPVS 007 are given in \S\ref{sec:results}. This section includes the best-fitting values of the physical parameters (ionization parameter, density, partial-covering parameter, and scaled column density). Derived parameters including column density, radial distance to the outflow, mass-outflow rate, and kinetic luminosity are given in \S\ref{sec:other}. Lastly, {in \S\ref{sec:discuss}, we revisit the \citet{Li2019} conclusions and} present our interpretation of the cause of the outflow variability from the results of our analysis. Throughout this paper, we adopt a flat universe $\mathrm{\Lambda-CDM}$ cosmology with $\mathrm{H_0 = 67.7\;km\;s^{-1}\;Mpc^{-1}}$, $\mathrm{\Omega_M = 0.31}$, and $\mathrm{\Omega_{\Lambda} = 0.69}$ \citep{2020Planck}.

\section{Observations and Data} \label{sec:obs}

This work primarily uses UV spectroscopy to investigate the underlying causes behind the clear changes in absorption-line structure of WPVS~007 over 7 years in the observed frame.  The interpretation of these changes is aided by the consideration of the continuum changes probed by extensive UV and optical \swift\ photometry. WPVS 007 has been regularly monitored with \swift\ from {2005} to the present day. Further details of the UV spectral and UV and optical photometric observations are given below. 

\subsection{UV Spectral Observations} 
WPVS 007 has been observed with UV COS spectroscopy with HST five times. The details of the observations including instrument, grating, date, wavelength range, spectral resolution, exposure time, and S/N at 1450~\AA\ are provided in Table~\ref{table:obs}. UV spectral observations with HST FOS in 1996 and FUSE in 2003 exist in the archive but were not included because they did not have comparable S/N (in the case of HST FOS) or wavelength range (in the case of FUSE) for a comparative analysis.  Interestingly, there is no evidence for the UV BALs in 1996, though the narrow, low-velocity mini-BAL is still evident in the spectrum \citep{2009Leighly}.

\begin{deluxetable*}{ccccccc}
    \tablecaption{ List of available UV HST COS observations of WPVS 007 \label{table:obs}}
    \tablehead{\colhead{Instrument} & \colhead{Grating} &\colhead{{Resolving}} &\colhead{Observation} & \colhead{Spectral} & \colhead{Exposure} & \colhead{S/N} \\[-2.0ex]
    \colhead{} & \colhead{} & \colhead{{Power}} & \colhead{Date} & \colhead{Coverage (\AA)} & \colhead{Time (s)} & \colhead{at 1450\AA}}
    \startdata 
    COS & G140L &1500-4000 & Jun 2010 & 1165.2--{1138.9\tablenotemark{a}} & 5061 & 23.5\\
        &       &          &          & {1230.2--2476.1} &      &     \\
    COS & G140L &1500-4000 & Jun 2013 & 1026.9--2496.4 & 4607 & 29.1\\
    COS & G140L &1500-4000 & Dec 2013 & 1026.7--2496.1 & 4607 & 21.1\\
    COS & G140L &1500-4000 & Mar 2015 & 1028.9--2338.8 & 5323 & 10.5\\
    COS & G140L &1500-4000 & Mar 2017 & 1028.9--2338.7 & 5495 & 13.9\\
    \enddata
    \tablenotetext{a}{{Two ranges due to detector gap.}}
\end{deluxetable*}

The final data products for all epochs were downloaded from the archive in January 2020 for analysis. No recalibration was necessary. Spectra were adjusted to rest-frame wavelengths using a redshift of 0.02861 adopting the NASA/IPAC Extragalactic Database redshift value.

The objective was to accurately model the shape of the available UV absorption lines in order to determine the physical properties of the outflow that cause them.  Modelling required that features have sufficient S/N.  We used the spectra in the 1065--1600~\AA\ range to include the \PV$\lambda\lambda1118,1128$ and the \CIII$\lambda1175$ multiplet BALs at the blue end of the wavelength range and the \CIV$\lambda\lambda1548,1551$ line at the red end of the wavelength range. The C{\sc iii}] complex at 1900~\AA\ has low S/N, and so we did not include this in our analysis.

Strong geocoronal lines are present from 1173--1189~\AA\ and from 1257--1273~\AA. We masked these sections from the analysis. We identified possible Galactic absorption lines from \CII\ and \SiII. Candidate Galactic lines were identified as narrow absorption lines that showed no change in optical depth between epochs. The region around the \CII$\lambda1334$ line was masked in all epochs. We identified Galactic lines of \SiII\ overlapping the \CIII\ trough, most obvious in the observation from 2015 when the \CIII\ BAL was weak. These  lines are likely \SiII$\lambda1190$ and $\lambda1193$ that have been shifted to 1157\AA\ and 1161\AA\ respectively. Because these latter features overlapped with BALs, we included them in the modeling rather than masking them out. {There is a detector gap in the 2010 epoch from 1139\AA-1230\AA\ in the rest frame.} 

\subsection{\swift\ Photometry} \label{sec:phot}
\swift\ has observed WPVS 007 since October {2005} typically with a cadence of once per week \citet{2013Grupe}. Observations prior to April 2013 are listed in \citet{2008Grupe,2013Grupe}.  The \swift\ X-ray telescope \citep[XRT;][]{burrows05} was operating in photon-counting mode \citep{hill04} and the data were reduced by the task {\it xrtpipeline} version 0.12.6., which is included in the HEASOFT package. Source counts were selected in a circle with a radius of 24.8$^{''}$ and background counts in a nearby circular region with a radius of 247.5$^{''}$. Due to the extremely low {X-ray} count rate of WPVS 007 of about $1\times 10^{-4}$ it cannot be detected in a single observation. A {X-ray} detection of WPVS 007 with \swift\ requires stacking the data of several years together. The UV-optical telescope \citep[UVOT;][]{roming05} data of each segment were coadded in each filter with the UVOT task {\it uvotimsum}. Source counts in all 6 UVOT filters were selected in a circle with a radius of 5$^{''}$ and background counts in a nearby source-free region with a radius of 20$^{''}$. UVOT magnitudes and fluxes were measured with the task {\it uvotsource} based on the most recent UVOT calibration as described in \citet{poole08} and \citet{breeveld10}. The UVOT data were corrected for Galactic reddening \citep[$E_{\rm B-V}=0.012$; ][]{sfd98}. The correction factor in each  filter was calculated with equation (2) in \citet{roming09} who used the standard reddening correction curves by \citet{cardelli89}. Due to new UVOT calibrations files it was necessary to re-analyze all previously published data.

\begin{figure*}[ht!]
\begin{center}
\includegraphics[scale=0.45,angle=0]{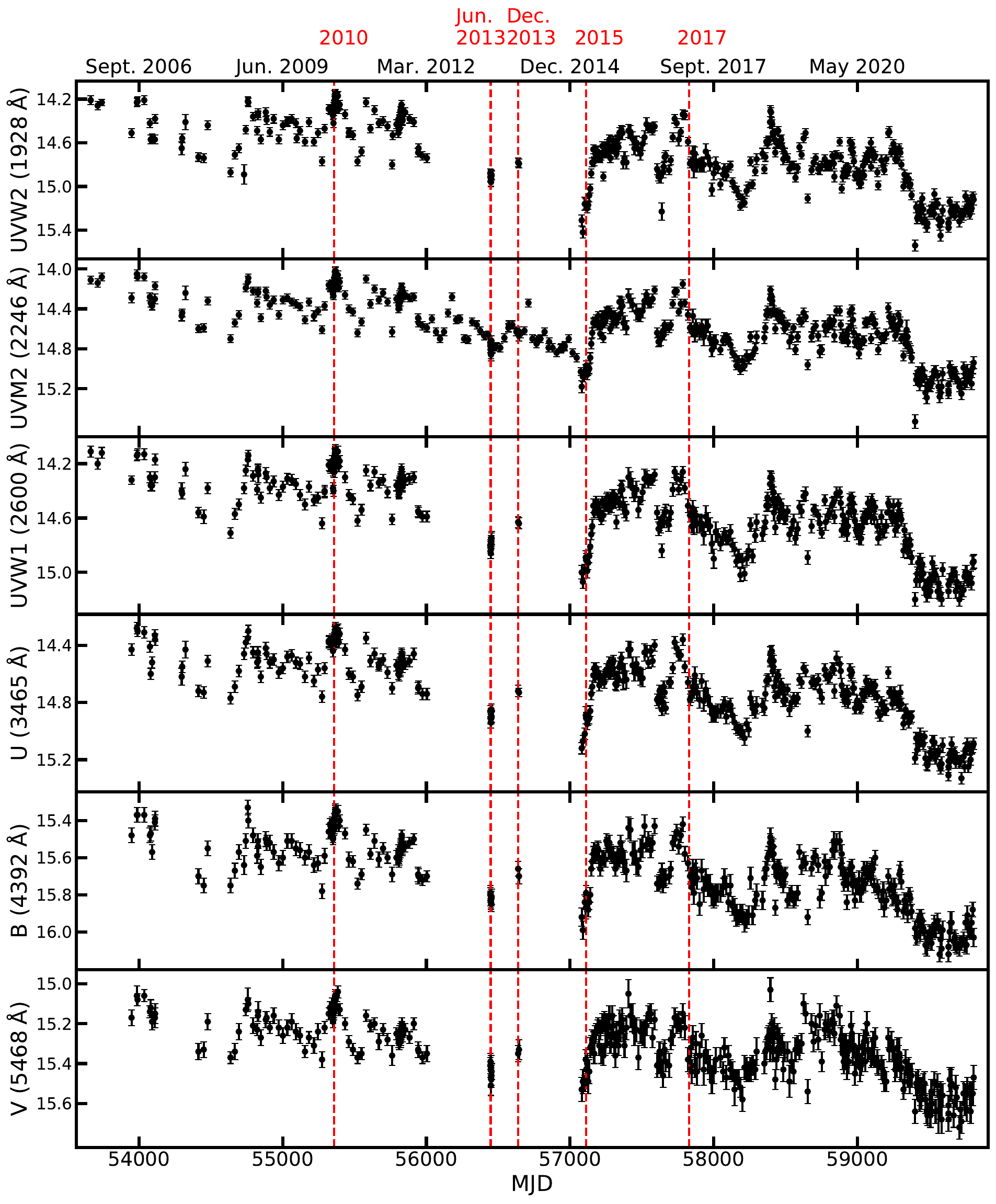}
\caption{All available \swift\ UV photometry over 16 years of monitoring WPVS 007. The red lines indicate the time of the HST COS observations. WPVS 007 is brightest in UV in 2010 and faintest in 2015. \label{fig:photo}}
\end{center}
\end{figure*}

We are providing updated photometry from continued \swift\ monitoring since April 2013 \citep{2013Grupe}. This observing campaign is the only regular monitoring of such an unusual low-luminosity source with a fully developed BAL. In Figure~\ref{fig:photo}, we show the results of the \swift\ monitoring campaign carried out over a period of 16 years. The \swift\ UVOT fluxes were corrected for Galactic extinction. As shown in \citet{Leighly2015}, the 2015 occultation event resulting in the {near} disappearance of the BAL shows a corresponding minimum in UV flux. The 2010 BAL was found when WPVS 007 was in a high state at the brightest magnitude. 

\begin{figure*}[ht!]
\begin{center}
\includegraphics[width=15cm]{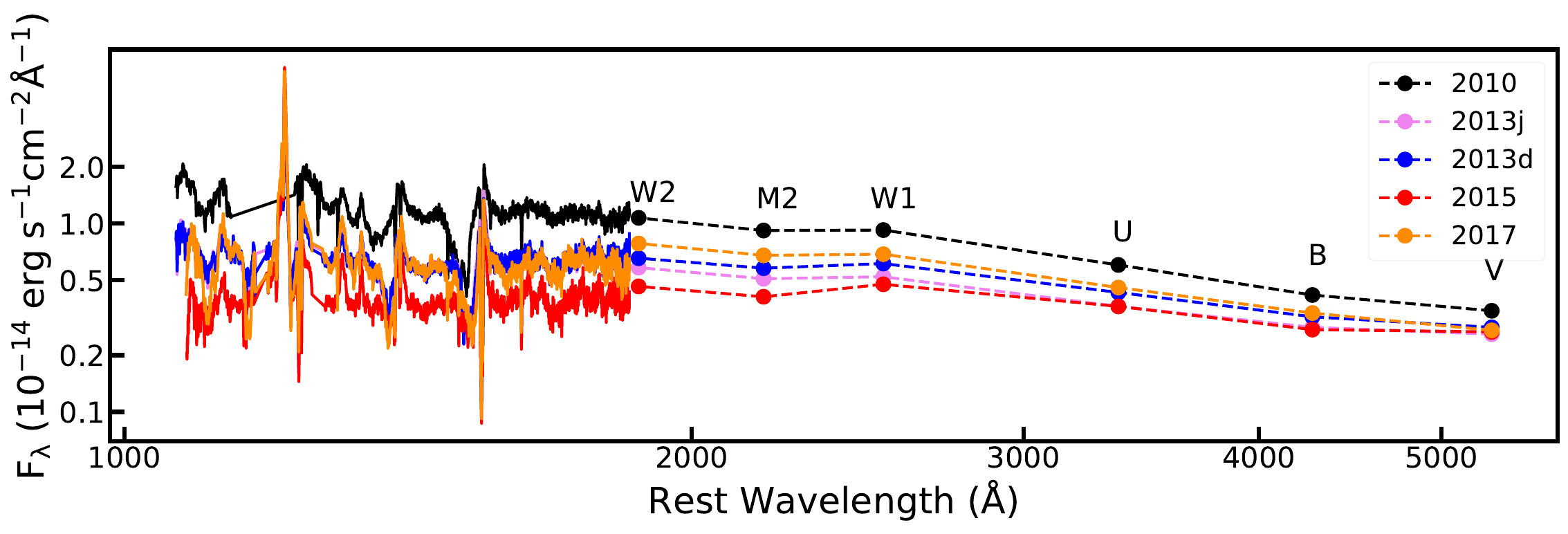}
\caption{HST COS observations and corresponding \swift\ SED for each epoch. {Errorbars for the Swift photometry are included but are smaller than the size of the plot points. Both \citet{Leighly2015} and \citet{Li2019} present similar plots (Fig. 2 in \citet{Leighly2015} and Fig. 2 in \citet{Li2019})\label{fig:sed_uv}}}.
\end{center}
\end{figure*}

Figure~\ref{fig:sed_uv} shows the SED using observations taken near the time of HST COS spectral observations, and is an updated version of the left panel of Figure 2 from \citet{Leighly2015}. Although the SED for 2010 (where the BAL was at maximum absorption) shows higher flux and 2015 (during the occultation event) shows reduced flux, the SEDs for December 2013 and 2017 with variable BALs are nearly identical. These observed changes are in contrast to the June 2013 and December 2013 epochs that show variation in the SED with almost no change in the BALs.

\begin{figure*}[ht!]
\begin{center}
\includegraphics[scale=0.4,angle=0]{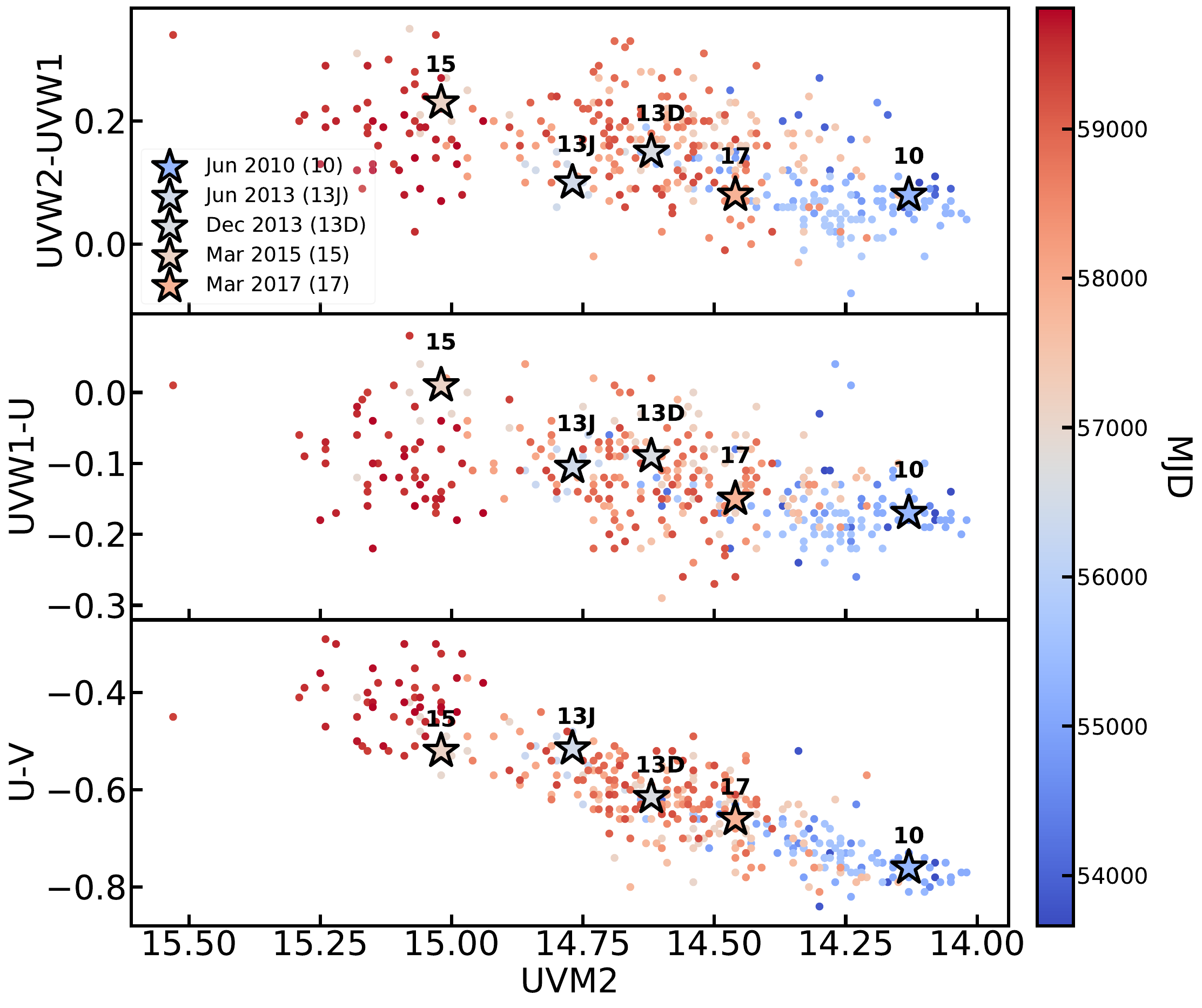}
\caption{Color-magnitude diagram of all available \swift\ photometry including the most recent observations. Colors represent MJD and stars represent the nearest \swift\ observations to the HST COS UV spectral observations. The central wavelengths of the UVW2, UVW1, U, and V bands are 1928\AA, 2600\AA, 3465\AA, and 5468\AA\ respectively. \label{fig:color_mag}}
\end{center}
\end{figure*}

Figure~\ref{fig:color_mag} shows the evolution of WPVS 007 in UV color-magnitude space, an updated plot from Figure 1 in \citet{Leighly2015}. There is a clear correlation between color and magnitude. The 2010 and 2015 observations were taken when WPVS 007 occupied 2 extremes in the color-magnitude diagram with the NLS1 appearing bluest in 2010 and reddest in 2015. The 2013 and 2017 colors are intermediate. {In Figure 1 of \citet{Leighly2015}, the authors showed that variable reddening is a better fit to the observed changes in color and brightness observed from the \swift\ photometry than intrinsic optical/UV spectral variability.}

\section{Spectral Modeling and Analysis} \label{sec:methods}

The spectra for each epoch were first fit using Sherpa for initial values of the continuum normalization and power-law indices, and the fluxes and FWHM values for the emission lines.  These were then fed into \simbal\ to enable fitting the absorption-line features for individual spectra.  The absorption features in many cases overlap with the emission lines.  Rather than mask out absorption features (with the exception of those mentioned in the previous section), absorption lines were fit with a series of Gaussians.  We used a power law to model the continuum and a series of Lorentzian and Gaussian lines to model the emission lines.  Based on prior experience, the UV resonance lines were initially modeled with two Lorentzian lines. As an exception, the \Lyalpha\ line was modeled with two Gaussian components, with one at $1215$\AA\ and a blue Gaussian component with the position, FWHM, and flux freed. We included an additional Gaussian component to model a blue wing of \CIV\ after finding that absorption models favored an additional emission component. Additionally, we included the spline fit model of weak \FeII\ lines at wavelengths greater than $\mathrm{1278}$\AA\ used in \citet{Leighly2015}. The full continuum model used in the \simbal\ analysis consists of a power-law component with the normalization and slope of the power law fit as parameters, the \FeII\ template used in \citet{Leighly2015}, and the emission-line model described above.  

A key factor to determine is whether the absorber covers the continuum and emission lines fully, partially, or not at all. The model that performed best in preliminary fits of the data was a model where the BAL does not cover the line-emitting region, but the mini-BAL covers both the line-emitting region and power-law continuum equally. This model indicates the presence of unabsorbed line emission through the BAL outflow. Other studies of BAL sources have found absorbers that do not fully cover the emission-line region \citep[e.g.,][]{1999Arav,2012Borguet,2022Choi}. 

\subsection{\simbal} \label{sec:simbal}
\simbal\ \citep{Leighly2018} is a novel forward-modelling and spectral-synthesis software package that has been used to model quasars with broad overlapping absorption troughs, leading for example to the discovery of the most powerful quasar outflow to date \citep{Choi2020}. \simbal\ uses grids of ion column density created using \textit{Cloudy} \citep{2017Ferland} to produce a synthetic spectrum from available atomic data and MCMC python package \verb|emcee| \citep{2013emcee} to step through the parameter grids.  

To generate the synthetic spectrum at each step, \simbal\ uses the ionization parameter, density, covering factor, and hydrogen column density. The column density is parameterized as $\log(N_{\rm H})-\log(U)$; here $U$ is the unitless ionization parameter defined as $ U=\frac{Q}{4\pi R^2 n_{\rm H} c}$, where $Q$ is the number of ionizing photons per second, $R$ is the distance of the BAL from the SMBH, $n_{\rm H}$ is the hydrogen density and $c$ is the speed of light. {Partial covering is now widely accepted as endemic for BAL absorption lines \citep[e.g.,][]{1997Barlow,1999Arav}, and can manifest as non-black saturation of BALs. Often this absorption is modeled as a step-function in column density, where the absorber covers part of the continuum uniformly, but light from the remaining uncovered region comes through unobstructed}. \simbal\ uses a power-law partial-covering model \citep[see][]{2002deKool,2005Arav,2005Sabra,Leighly2019} where $\tau=\tau_{\rm max}x^a$ {with $x$ as the projection of the 2-D region onto 1-D with a value between 0 and 1 and $\log(a)$ is the parameter fit by \simbal. Power-law partial covering assumes a power-law distribution of optical depth, where the power-law index is a fit parameter for \simbal. When the index value is 0, this is equivalent to the homogeneous partial covering case, e.g., a sharp-edged, solid absorber. \simbal\ uses power-law partial covering because it is computationally tractable and explains the observed difference in covering fraction between high and low ionization lines \citep[e.g.,][]{2001Hamann}.} See \citet{Leighly2018} for further details about \simbal\ and the spectral synthesis method of modelling BAL quasar absorption and \citet{Leighly2019} for further discussion of the covering fraction.

{An advantage of \simbal\ over traditional analysis methods is that the solution is self consistently constrained by all of the lines simultaneously, as well as from the upper limits of the absorption lines that are not detected. Therefore, in a velocity-resolved analysis (such as the analysis presented here), the physical parameters of the gas are constrained by multiple lines. This allows for excellent constraints on physical conditions of the gas due to the diagnostic power of lines with different physical properties such as oscillator strengths and abundances. The first velocity-resolved analysis using \simbal\ was the of low-ionization BAL source SDSS J$085053.12+445122.5$ \citep{Leighly2018,Leighly2019}. Applied to multi-epoch data, this methodology offers the capacity to study the change in the physical outflow properties over time as a function of velocity.}

\subsection{Absorption}

\begin{figure*}[ht!]
\begin{center}
\includegraphics[width=15cm]{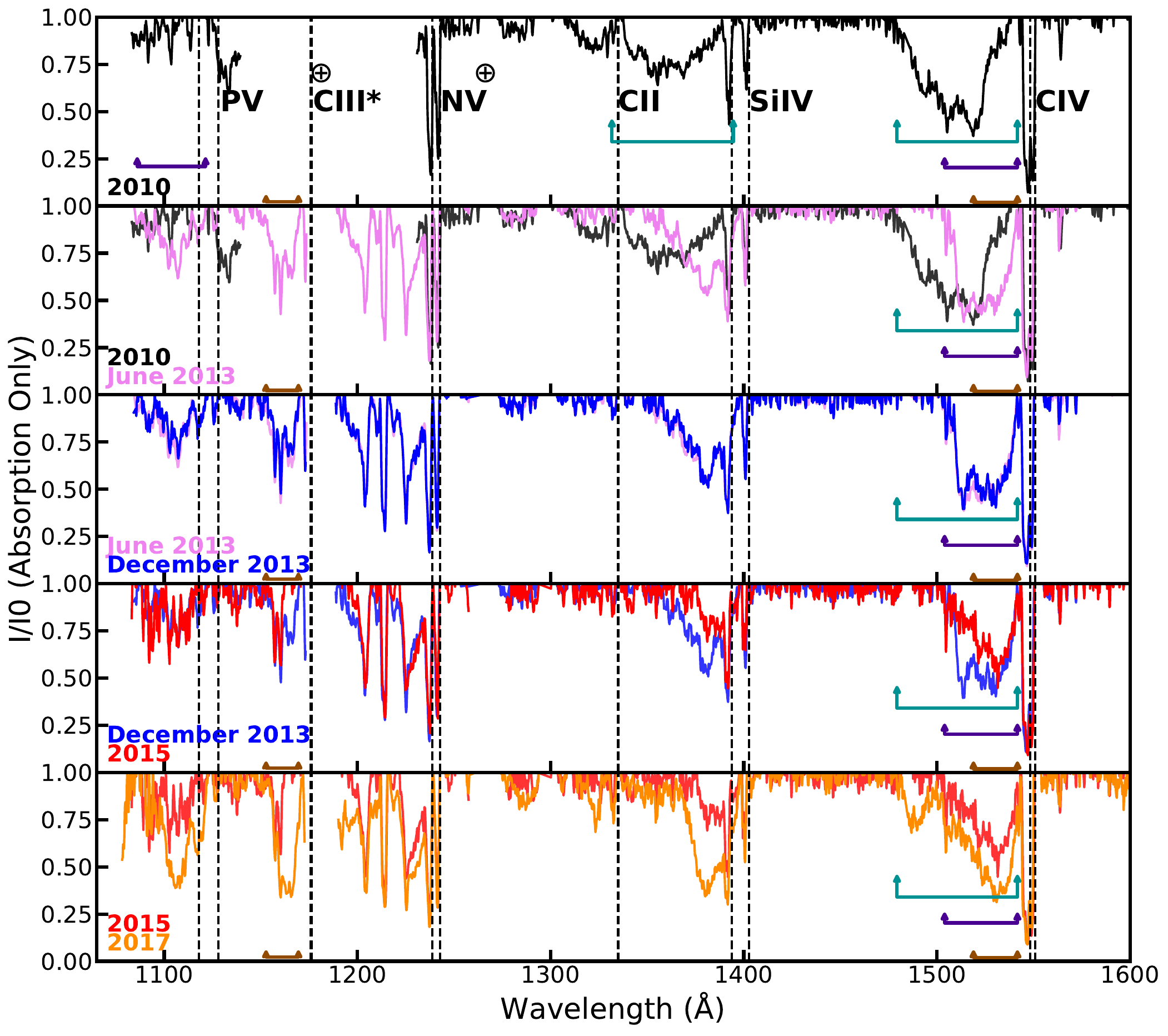}
\caption{Continuum-normalized HST COS spectra showing the strong variability of high-ionization BALs between epochs of observation. Major lines are labelled and locations of geocoronal lines where data has been masked are indicated with the $\Earth$ symbol. The teal bar shows a velocity range of $\sim-13,000$ \kms, the purple bar shows a velocity range of $\sim-8,500$ \kms, and the brown bar shows a velocity range of $\sim-5,000$ \kms; these bars indicate the velocity ranges of interest as described in \S\ref{sec:methods}. \label{fig:asbvar}}
\end{center}
\end{figure*}

In Table~\ref{table:emlines}, we list key lines modeled in absorption, including their ionization potential and primary diagnostic power for determining physical conditions of the gas. Figure~\ref{fig:asbvar} shows the differences in absorption between epochs after continuum normalization.  To find a generic absorption model that would work for all epochs, we first look at the overall absorption structure in all epochs and at which BALs and BAL sub-features change over time. Subsequently, a starting model for absorption lines was chosen taking into consideration the morphology of the lines, total width of the absorption lines in all epochs, and variation within the structure of individual BAL lines between epochs. 

The ionizing continuum used in the \textit{Cloudy} grids is a relatively soft quasar SED \citep{2011Hamann}, that we use because WPVS 007 has been consistently found to be X-ray weak, with a soft X-ray spectrum \citep{1995Grupe,2007Grupe,2008Grupe,2013Grupe}. We also assumed an enhanced metallicity $({Z=3Z_{\odot})}$ consistent with previous findings that BALQs have metallicities $ Z\gtrsim Z_{\odot}$ as measured using emission-line ratios (\citealt{2002Hamann}, but see also \citealt{2021Temple}).

\begin{deluxetable*}{cccc}
    \tablecaption{ Summary of Key Spectral Line Diagnostics \label{table:emlines}}
    \tablehead{\colhead{Ion} & \colhead{Wavelength (\AA)} & \colhead{IP (eV)\tablenotemark{a}} & \colhead{Diagnostic}}
    \startdata 
    \PV & 1118, 1128 & 65 & Column Density\tablenotemark{b} and Ionization Parameter\tablenotemark{c}\\ 
    \CIII & 1175\tablenotemark{d} & 48 & Density\tablenotemark{e} and Column Density\\
    \NV & 1239, 1243 & 98 & Ionization Parameter\\
    \CII & 1335, 1336 & 24 & Column Density\tablenotemark{f}\\
    \SiIV & 1394, 1403     & 45 & Covering Fraction and Ionization Parameter\\
    \CIV & 1548, 1551    & 64 & Covering Fraction and Ionization Parameter\\
    \enddata
    \tablenotetext{a}{Ionization Potential in units of eV}
    \tablenotetext{b}{ \citet{1998Hamann}}
    \tablenotetext{c}{ \citet{Leighly2018}}
    \tablenotetext{d}{ Multiplet of 5 lines, each is modelled as a separate line in \simbal.}
    \tablenotetext{e}{ \citet{Gabel2005}}
    \tablenotetext{f}{ \citet{2019Hazlett}}
\end{deluxetable*}

Of the physical outflow parameters listed in Table~\ref{table:emlines}, density is the hardest to constrain with only the \CIII\ line. {In order to constrain density, we need absorption from the collisionally excited state of an ion \cite[e.g.,][]{Gabel2005,2013Arav}. In this case, \CIII\ is the only ion within this wavelength range producing absorption from an excited state.} Lines with high ionization potential, such as \NV, are sensitive to changes in ionization parameter. As was discussed in \citet{Leighly2018}, the \PV\ line is very useful as a constraint on column density {because it is rarely optically thick due to the low abundance of \PV\ }\citep{1998Hamann} and also constrains the ionization parameter {as a high ionization is required to produce \PV\ }\citep[see also][]{2009Leighly}. We observe the \PV\ line in all COS spectra, which is an indication that the much more abundant \CIV\ will likely be saturated \citep{2012Borguet}. 

{The kinematic differences and opacity differences (in the case of \CIII) required us to look for ways to tie or fix parameters in the model.} As a first step, we look for ways to group structures within the BAL.  Based on relative velocity widths of various lines, we divided the velocity profile into {4} main regions, a high-velocity gas component as all absorption between $\sim$--12,000 \kms\ and $\sim$--8,000 \kms, a medium-velocity component consisting of absorption between $-8,000$ and $-5,000$ \kms, and {two} low-velocity component{s} with absorption between {$\sim-5,000$ \kms--$\sim-3,000$ \kms and $\sim-3,000$ \kms--$\sim-600$ \kms, which we call low (1) and low (2) respectively}. The \CIII\ line (that we rely on to constrain the density; see Table~\ref{table:emlines}) absorbs only at low velocities ($\sim-$5,000 \kms--$\sim-$600 \kms) in all epochs{, but due to differences in absorption between low (1) and low (2) in the 2015 epoch compared to other epochs, we subdivided this velocity group further}. The high-velocity gas is only present for some high-ionization lines such as \CIV\ and \NV\ in the 2010 and 2017 epochs, and not low-ionization lines such as \CIII\ or \CII, and is not present in \PV. The mini-BAL feature at velocities of $\sim-$600 \kms\ overlaps the \CIV\ emission line (with strong absorption for other high-ionization lines of \SiIV\ and \NV). The mini-BAL absorption line is deeper than the BAL and does not show variation between epochs. The observed lack of variation implies a larger covering fraction for the mini-BAL than the BAL. For the remainder of the paper, we discuss absorption variability as a function of the velocity groups defined. 

We used the tophat setting in \simbal\ \citep[see][]{Leighly2018,Leighly2019,Choi2020} to model the spectra. The tophat accordion model uses rectangular bins of equal velocity width to span the BAL. Both the maximum velocity and width per bin are fit as parameters in \simbal, but the number of bins remains fixed. The starting bin size was chosen to be $950\;$\kms, roughly half the separation width of the \SiIV\ doublet, or the velocity separation between the \NV\ lines. We used \CIV\ to select the number of bins necessary to model absorption across all epochs. 
{The number of bins for each epoch was chosen so that variability in each sub-feature of key BALs could be studied between epochs. Specifically, we wanted to ensure that the number of bins in each epoch resulted in the final velocities of each bin roughly aligned between epochs. We chose the smallest number of bins to achieve this objective for each epoch. }
Using the velocity groups defined above, we use 12 bins to span absorption from velocities of $-13,000$ \kms\ to $-1,200$ \kms\ in 2010. For the 2013 epochs, we used 7 bins to span velocities from $-8,000$ \kms\ to $-1,200$ \kms, the 2015 epoch was fit with 8 bins spanning from $-9,000$ \kms\ to $-1,200$ \kms, and the 2017 epoch was fit with 11 bins spanning velocities from $-12,000$ \kms\ to $-1,000$ \kms. We use two tophat bins to fit the mini-BAL absorption. These tophats had widths of 300 \kms\ spanning velocities from $-600$ \kms\ to 0 \kms. \citet{Leighly2018} looked at the difference in {final fit }parameters when varying the total number of bins that span the BAL and found that the total number of bins {used to model the absorption did not significantly affect the values of the final fit parameters.}

To reduce the number of free parameters, we tied some absorption parameters between \simbal\ bins within the same velocity group. Following the analysis of \citet{Leighly2018}, each bin has a unique value of ${\log(N_{\rm H})-\log(U)}$ and partial covering parameter $\log(a)$. Ionization parameter is tied between bins in a given velocity group {(with unique ionization parameters assigned to low (1) and low (2)) }to reduce the number of fit parameters (and therefore computation time). Ionization parameter is more difficult to constrain with the lines available in the spectrum, and we further found that freeing ionization parameter between bins did not improve the fit. With \CIII\ as the only density-sensitive line present in all spectra, we fit only one density value for each epoch. The low-velocity bins overlap with the \CIII\ line and are constraining the density value for this model. 
{We tested whether freeing the density between velocity groups improves the fit, but found that this failed an F-test for significance.}

\subsection{{Best-fit Results}}

\begin{figure*}[ht!]
\begin{center}
\includegraphics[width=16cm]{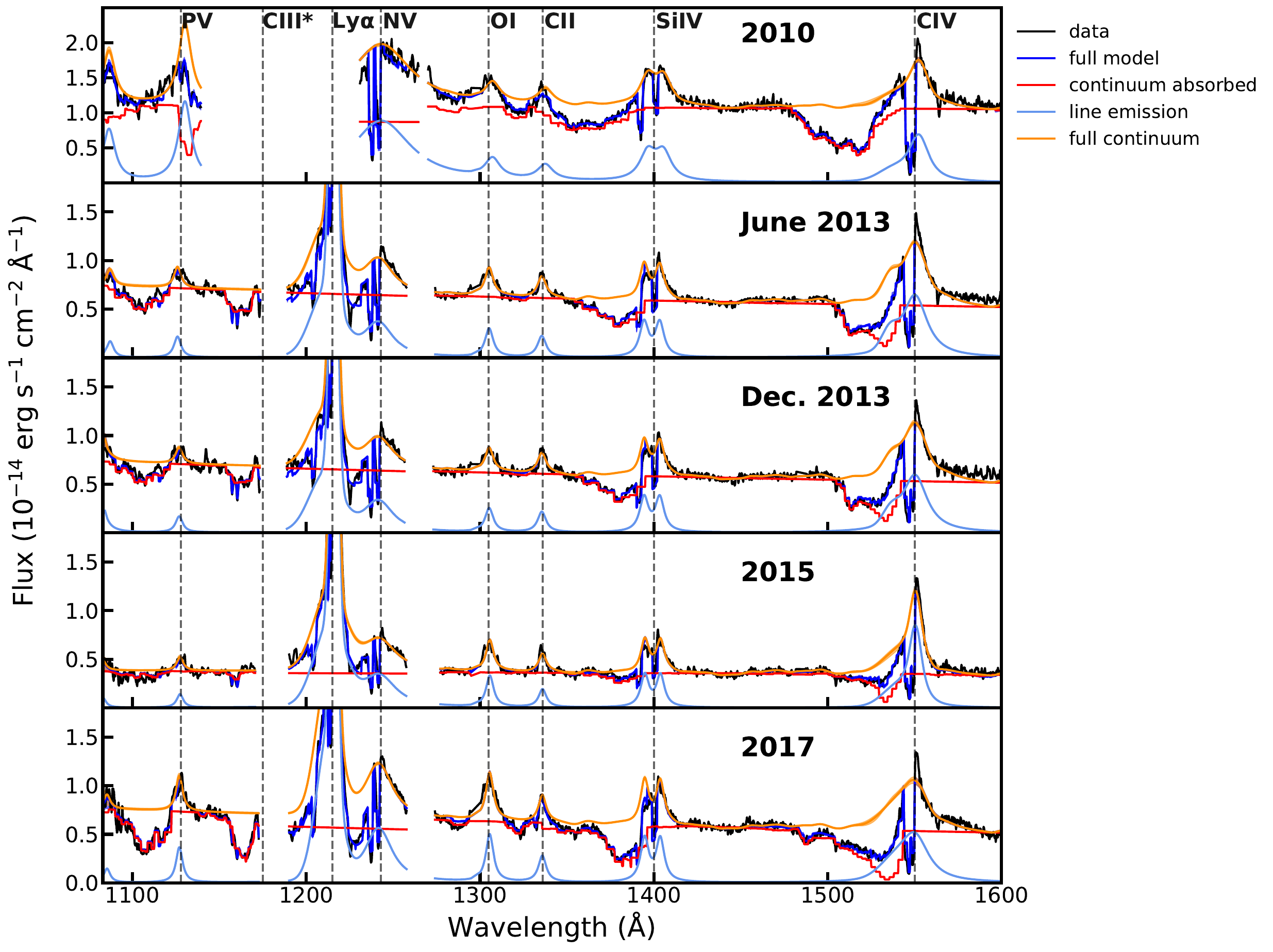}
\caption{Best-fit \simbal\ models of all five HST COS epochs fit independently. The full model is shown in blue, while we present the absorbed continuum model (absorption without emission lines or mini-BAL absorption) in red. The orange line shows the full continuum (power law and emission lines), and the light-blue line shows the line emission only. The gap in data in the 2010 epoch was due to the observing mode chosen, so the \CIII\ and \Lyalpha\ lines were not observed. Spectra were masked in all epochs between the \CIII\ and \Lyalpha\ line{ from 1173\AA\ to 1190\AA\ and between 1258\AA\ and 1274\AA\ in the rest frame} due to strong geocoronal line emission.\label{fig:singlefits}}
\end{center}
\end{figure*}

\begin{figure*}[ht!]
\begin{center}
\includegraphics[scale=0.28,angle=0]{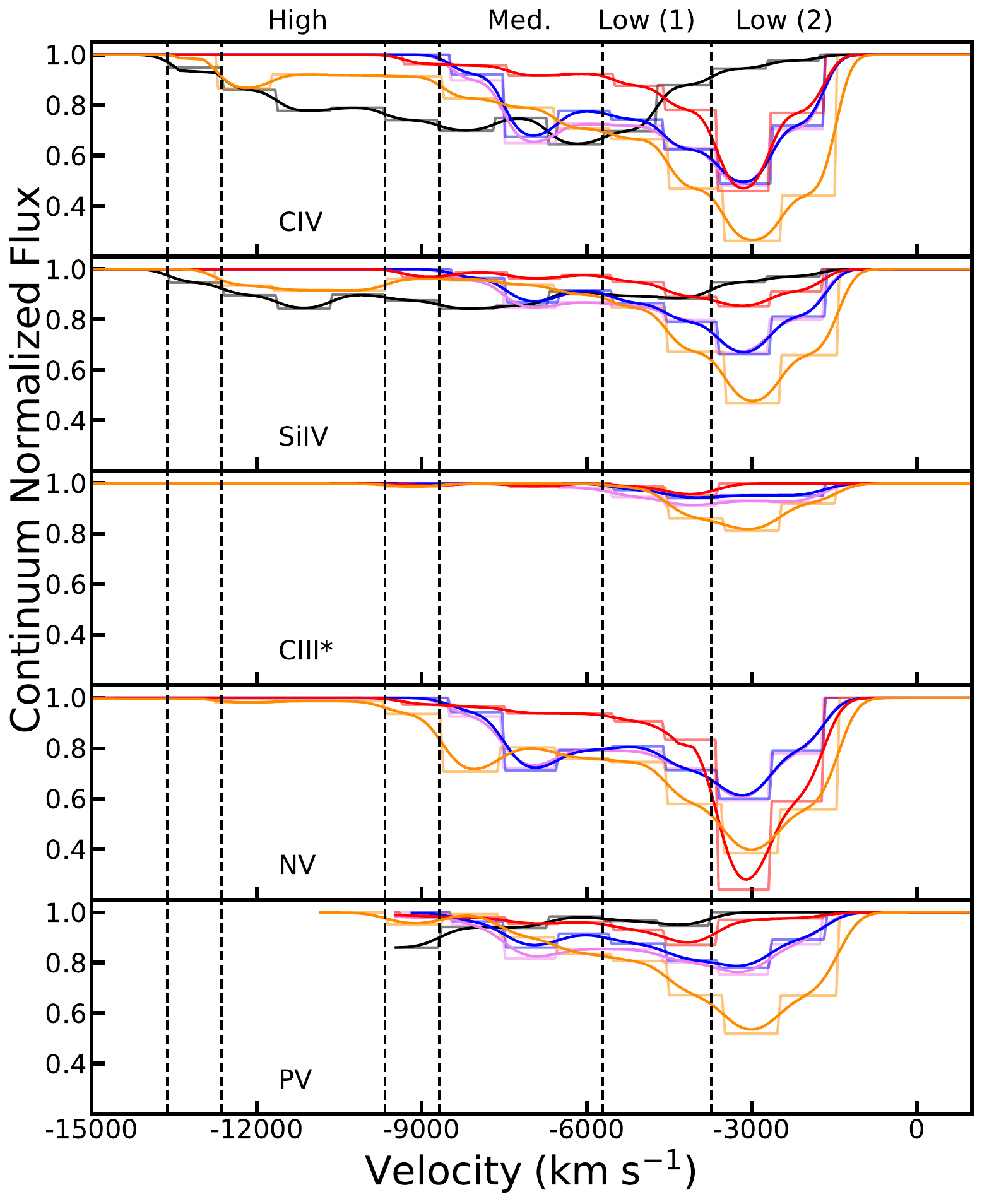}\hspace{1cm}\includegraphics[scale=0.28,angle=0]{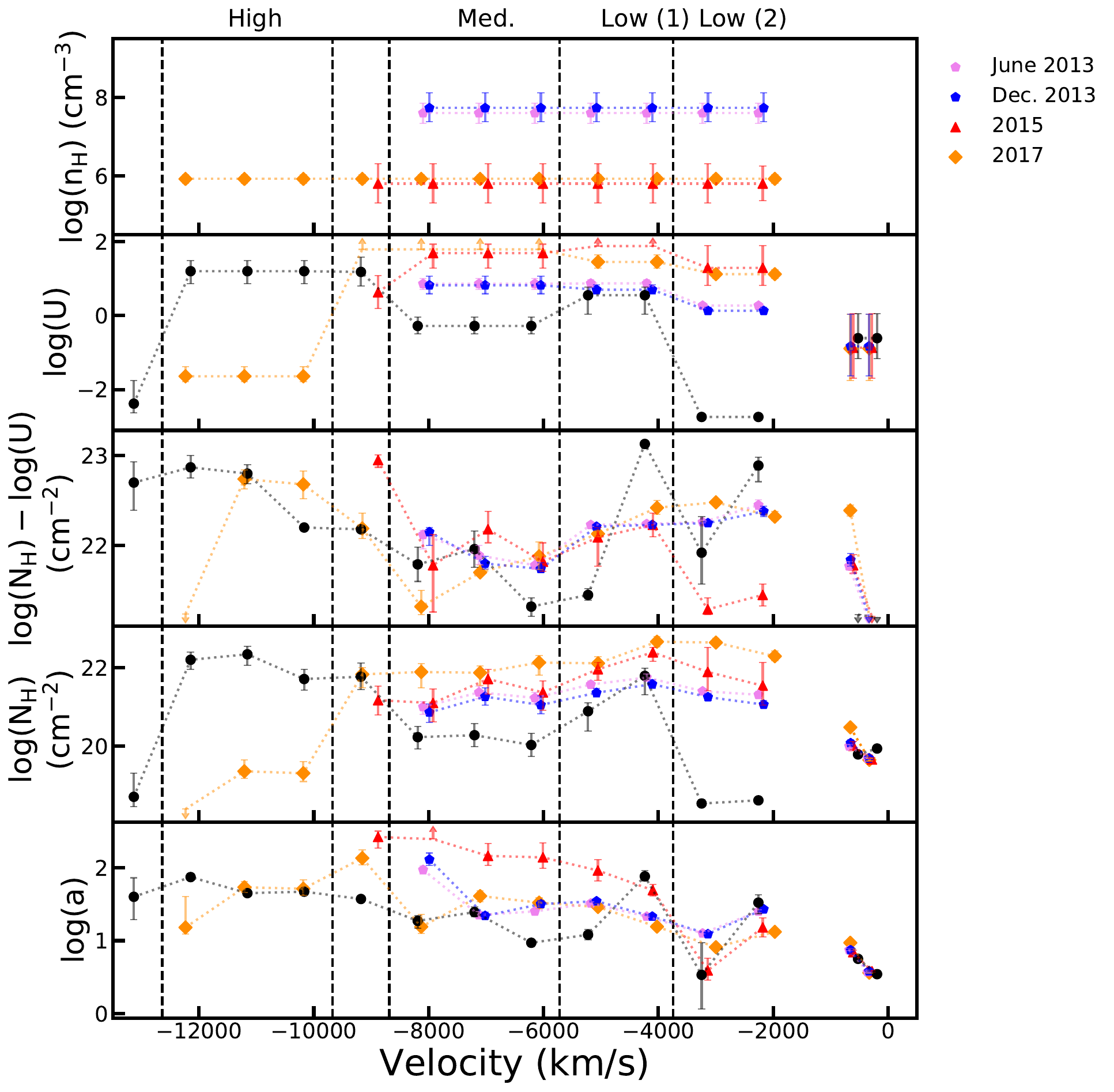}
\caption{(left) Absorption troughs as a function of velocity for each epoch for \CIV, \SiIV, \CIII, \NV, and \PV. The {stepped} lines show the \simbal\ models for each ion and the smooth curves show the same model smoothed with a gaussian function to help show the differences between epochs. Absorption from the mini-BAL is not shown. (right) Best-fit parameters as a function of velocity for all epochs. Values and errorbars correspond to median and 95\% confidence interval values sampled from the posterior parameter distributions of each parameter. Vertical dashed lines mark velocity groups that have the same ionization parameter. Points to the right of the parameter plot correspond to the mini-BAL parameters. \label{fig:singleepochmod}}
\end{center}
\end{figure*}

As a first step we fit each epoch independently, letting all outflow, continuum, and emission-line parameters vary between epochs. In Figures~\ref{fig:singlefits} and \ref{fig:singleepochmod}, we present the fits and best-fit parameters respectively for the BAL and mini-BAL for all epochs.{ We note that the continuum in the 2013 epochs shows an uneven blue wing for the \CIV\ line. We tested the impact of this shape on our final results by fitting a separate model for the December 2013 epoch with additional constraints to force a smooth shape for the final \CIV\ emission line. The resulting final parameters were consistent with the original fit as expected because the \CIV\ line is saturated, and therefore is not strongly constraining on the best-fitting model parameters.}

The physical parameters as a function of velocity are shown for each HST COS epoch fit in the right panel of Figure~\ref{fig:singleepochmod}. Density is not shown for the mini-BAL as it could not be constrained. The left panel of Figure~\ref{fig:singleepochmod} shows synthetic ion models produced by \simbal\ using the final best-fit models for each epoch as a function of velocity. Dashed lines are used to show the corresponding BAL features for each velocity group. Velocity bins between main velocity groups were given unique ionization parameter values and not tied. Although the density was fit for the 2010 epoch (and shown in Figure~\ref{fig:singleepochmod}), we do not present any parameters derived from the density value for 2010 as no density sensitive lines were present in the band pass.  {We note that the high-velocity component in 2017 has a lower ionization parameter than the low-velocity gas and yet shows no evidence of absorption from low-ionization line \CIII. The \CIII\ line is sensitive to more than ionization state alone. \simbal\ is fitting the column density, ionization parameter, and covering fraction simultaneously for this component, all of which contribute to the opacity produced (or not produced in the case of the high-velocity gas component) in the synthetic spectrum.}

\section{What Physical Parameters Drive BAL Variability{ in WPVS 007}?} \label{sec:results}

We used \simbal\ to  fit the continuum, emission-line and absorption line features in 5 epochs of HST UV spectra of the NLS1 WPVS 007.  The broad absorption lines {present in the spectra between 1090\AA\ and 1600\AA\ including} \PV, \CIII, \Lyalpha, \NV, \CII, \SiIV, and \CIV\ were fit simultaneously as a function of velocity.  Overall, our \simbal\ results from fitting the individual epochs independently were grossly consistent over time.  We found that the spectral fits are not very sensitive to changes in density and interchanging density values between epochs produces little change in the observed models. {This makes sense because only \CIII\ in our bandpass probes density. }We do not explore a change in density as a potential cause of the variability further. We were unable to pick out a clear single cause of the variability between a change in ionization parameter, column density, or partial covering from the results presented in Figure~\ref{fig:singleepochmod}. 

In the following sections, we turn to alternative methods for {determining} the primary cause of the variability including fitting more than one epoch simultaneously using \simbal\ and quantifying the observed depth of the individual BAL lines as a function of velocity to  investigate potential correlations between the depth of the BAL and the change in physical parameters.

\subsection{Multi-Epoch Fits}\label{sec:multiepoch}

We simultaneously fit 2 epochs at a time, {tying} all absorption parameters between epochs except one to see if a single parameter can explain the variability between epochs. These simultaneous fits of WPVS 007 are the first time more than one epoch of observation was fit simultaneously using \simbal. This analysis was done with the December 2013, 2015, and 2017 HST COS observations. The 2010 observation was excluded due to missing observations of \CIII\ and \NV\ lines, and we chose the December 2013 observation over the June 2013 observation as it is chronologically closer to the 2015 and 2017 observations. The 2013 and 2017 epochs were chosen first due to the comparable signal-to-noise and little change in continuum or line parameters between the two epochs. In the end, simultaneous fits of 2013--2017 (not including 2015), 2013--2015, and 2015--2017 epochs were carried out.

The scenarios tested are as follows: (1) there is a change in ionization state of the gas, which will be reflected as a change in ionization parameter ($U$); (2) there is a change in column density; or (3) there is a change in partial covering ($\log(a)$). As \simbal\ fits the column density with respect to the hydrogen ionization front (${\log(N_{\rm H})-\log(U)}$) instead of a total column density, we {tied} the column density for scenario (1) listed above. 

\begin{figure*}[ht!]
\begin{center}
\includegraphics[scale=0.35,angle=0]{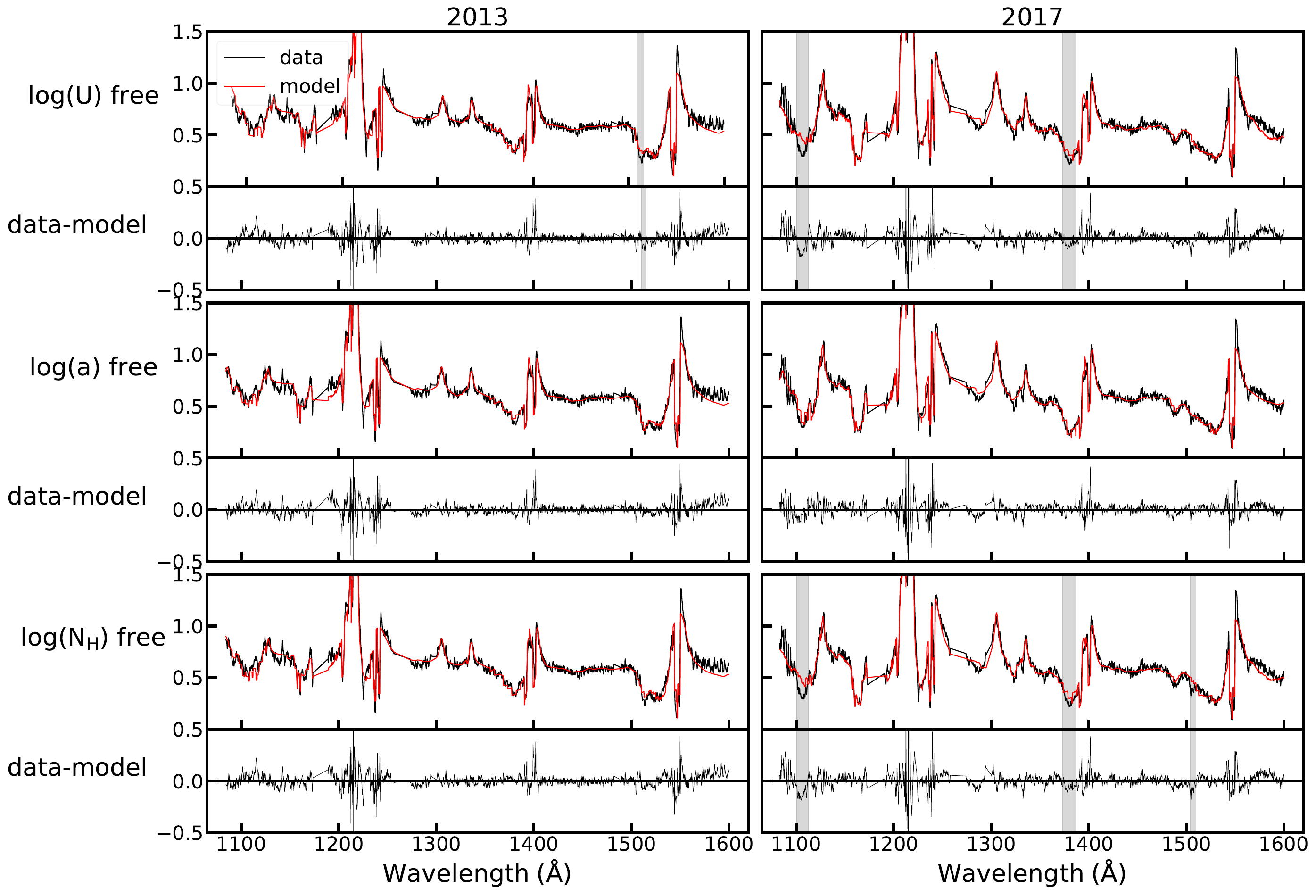}
\caption{Results of the simultaneous fits of the 2013 and 2017 epochs where only one parameter is allowed to vary between the two epochs. Regions highlighted in gray are regions where the model produces a worse fit than the single-epoch fit of the spectrum. By varying only {the }ionization parameter (top row) the \PV\ and \SiIV\ BALs are not fully modeled. When the column density is freed (bottom row) between epochs the \PV\ and \CIV\ BALs are not well fit. For the $\log(a)$ or covering fraction freed model (middle row), all lines for both epochs are well modeled.  \label{fig:multi_epoch_acxcomp}}
\end{center}
\end{figure*}

In Figure~\ref{fig:multi_epoch_acxcomp}, we show the results of the 2013 and 2017 simultaneous fitting for the three test cases outlined above. The models fail to fit the full depth of the \PV\ and \SiIV\ BALs in both scenario (1) where we assume that the ionization parameter is causing the variability and (2) where we assume that the column density alone is controlling the variability. In contrast, all absorption features of both epochs can be fit by varying the partial-covering parameter, $\log(a)$ only. The variation in absorption lines can be therefore explained as largely driven by a change in the amount of partial covering. 

Given that the $\log(a)$ parameter is the main driver for absorption-line variability between epochs, we further investigate if freeing either ionization parameter or column density in addition to partial covering improves the observed fit from the $\log(a)$ fits alone; these models fail the F-test for improvement to the model given the additional parameters.

{The changing covering fraction best explains the changes between 2013 and 2017 and is less satisfactory for the ultra low state in 2015.}
We test {whether} these results are consistent with the 2015 data by fitting the 2015 spectrum with the best-fit values of density, ionization parameter, and column density from the combined 2013-2017 fit and allowing only the continuum parameters, velocity, and $\log(a)$ parameters to vary. Using an F-test{,} we find that the fit is improved statistically when all absorption parameters {(not just $\log(a))$ }are allowed to vary. The residuals show that the difference between the models{ (where only $\log(a)$ is free vs. where all absorption parameters are free)} is mostly overlapping the emission lines and mini-BAL area, indicating that there may be either some small change in the mini-BAL in 2015 or that there is some degeneracy between the continuum parameters and the mini-BAL absorption parameters. 
As far as differences between the BAL models, the difference comes in a slightly poorer fit for \CIII\ in the two lowest velocity bins where we see the BAL disappear in 2015. The change in covering fraction is not able to explain the low opacity in 2015 for these lowest velocity bins, which may indicate that there are slight variations in other parameters at low velocities between these epochs that are not being taken into account when only $\log(a)$ is allowed to vary. 

\subsection{Correlation with Empirical Parameters}\label{sec:correlation}

We wanted to see if we can determine the cause of variability using the individual fits of each epoch to confirm the result we obtained with the simultaneous fits of the 2013 and 2017 epochs. To do this, we look for correlations between the change in depth of the BAL and the change in physical parameters from the final \simbal\ fits presented in Figure~\ref{fig:singleepochmod}. Throughout this analysis, we {used} a change in bin depth between epochs{ to measure the absorption strength of the BAL at a certain velocity}. As our velocity bins are $~\sim1000$ \kms, this change in bin depth is comparable to the statistic $\Delta A$ used by {\citet{Capellupo2011,Capellupo2012,Capellupo2013}} to quantify changes in BALs over time. {Their $\Delta A$ parameter is the change in the fraction of the continuum-normalized flux removed over 1000 to 2000 \kms\ velocity bins \citep{Capellupo2011}. }

\begin{figure*}[ht]
\begin{center}
\includegraphics[scale=0.27,angle=0]{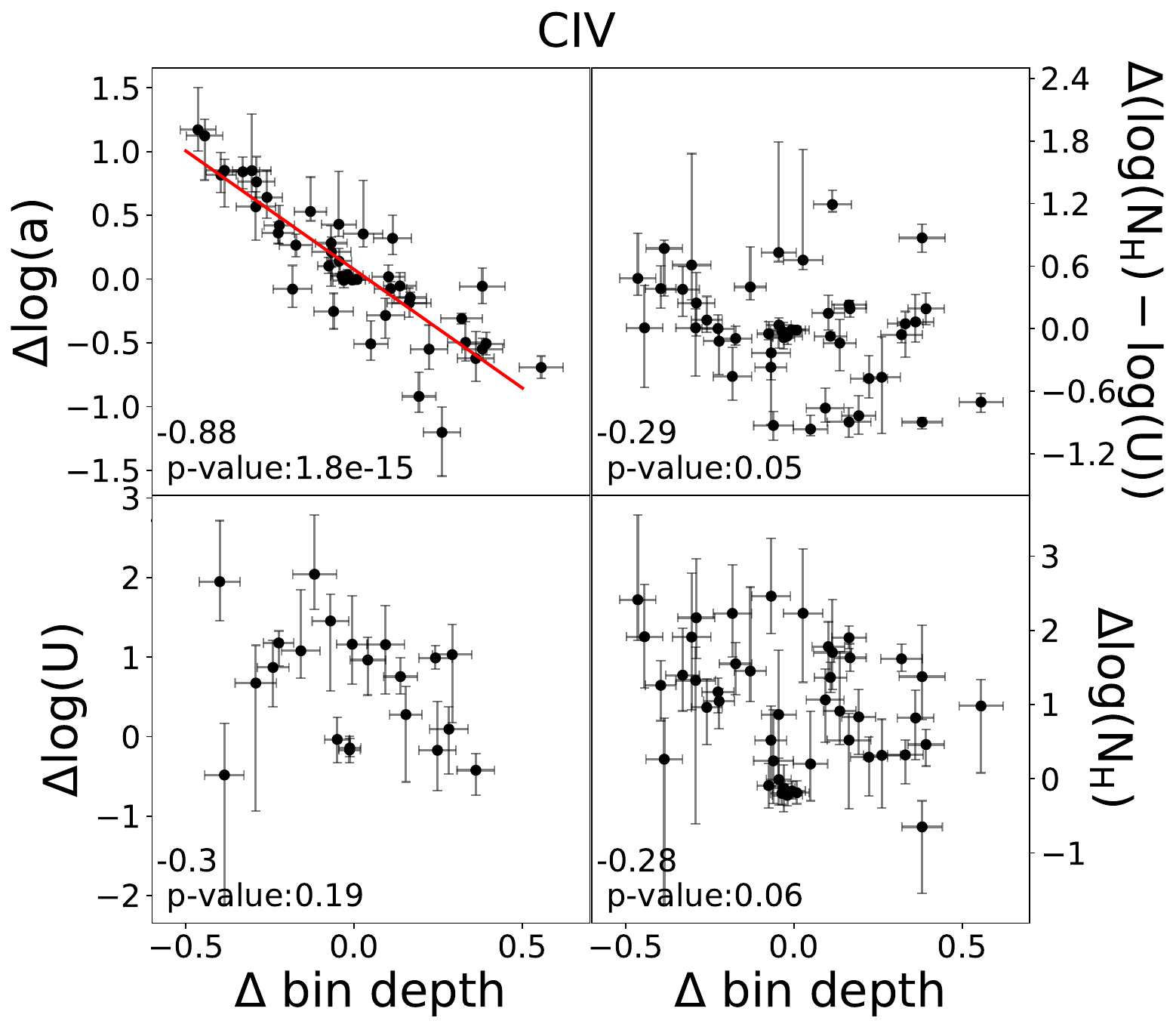} \includegraphics[scale=0.27,angle=0]{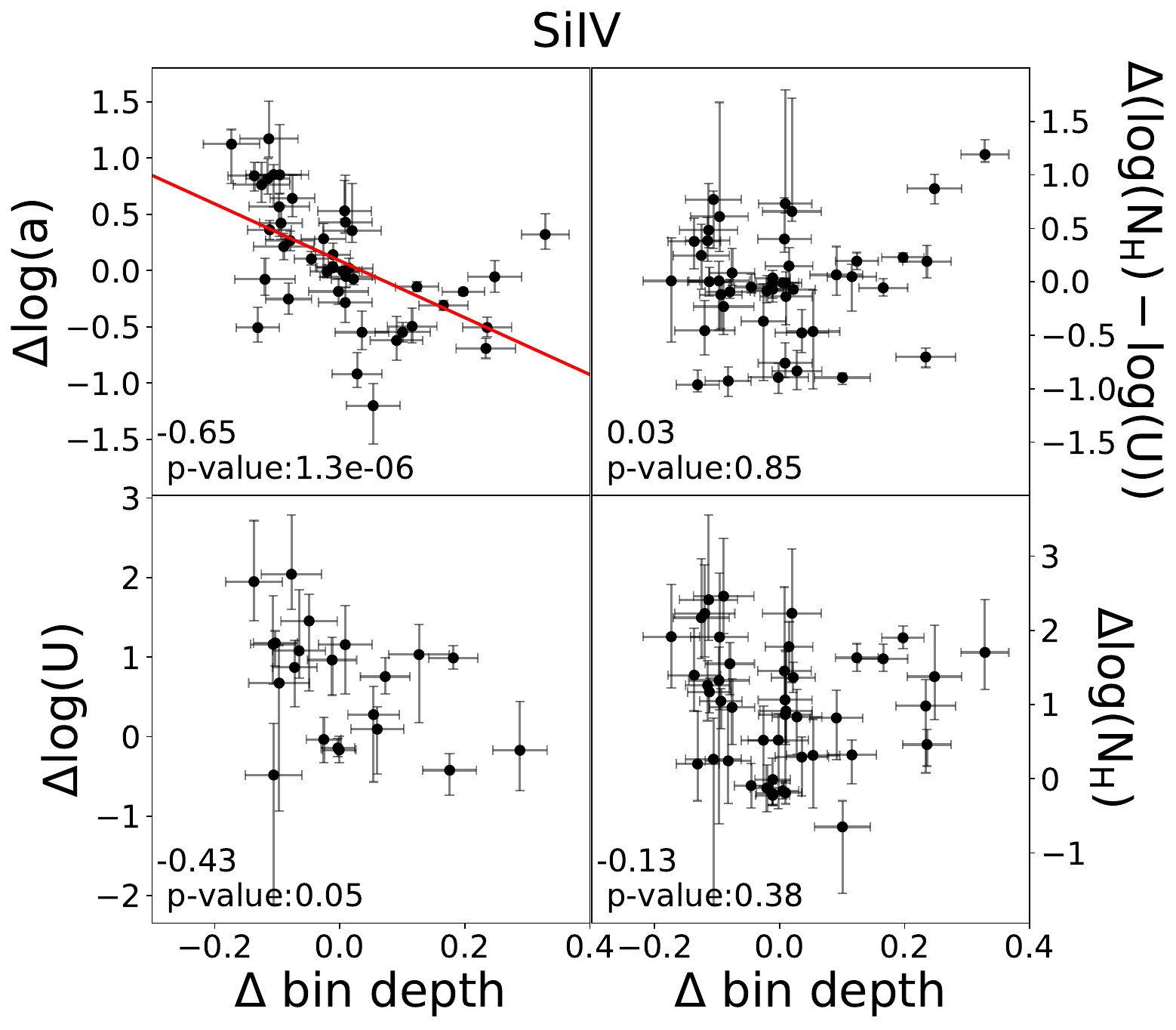} \\
\includegraphics[scale=0.27,angle=0]{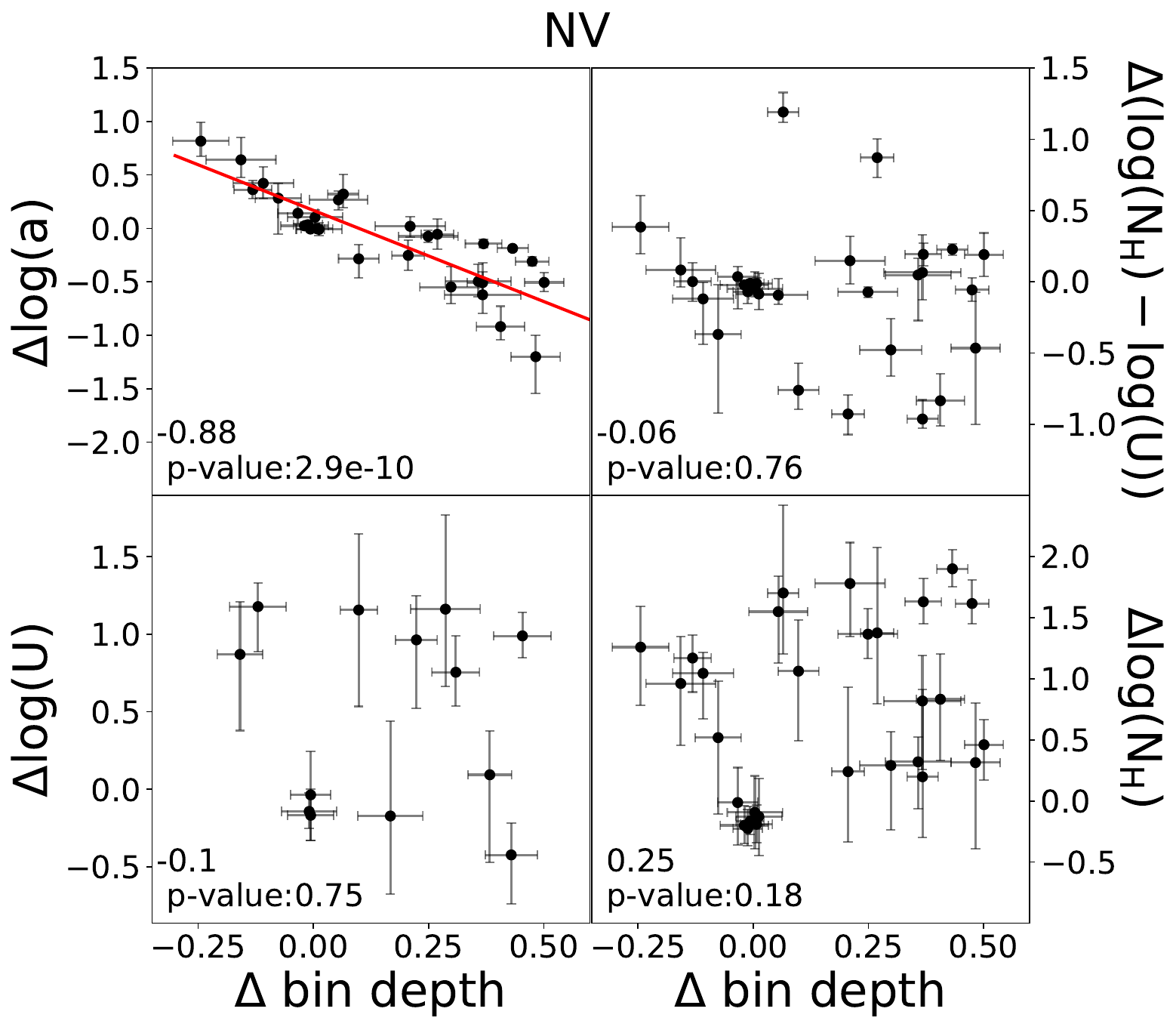} \includegraphics[scale=0.27,angle=0]{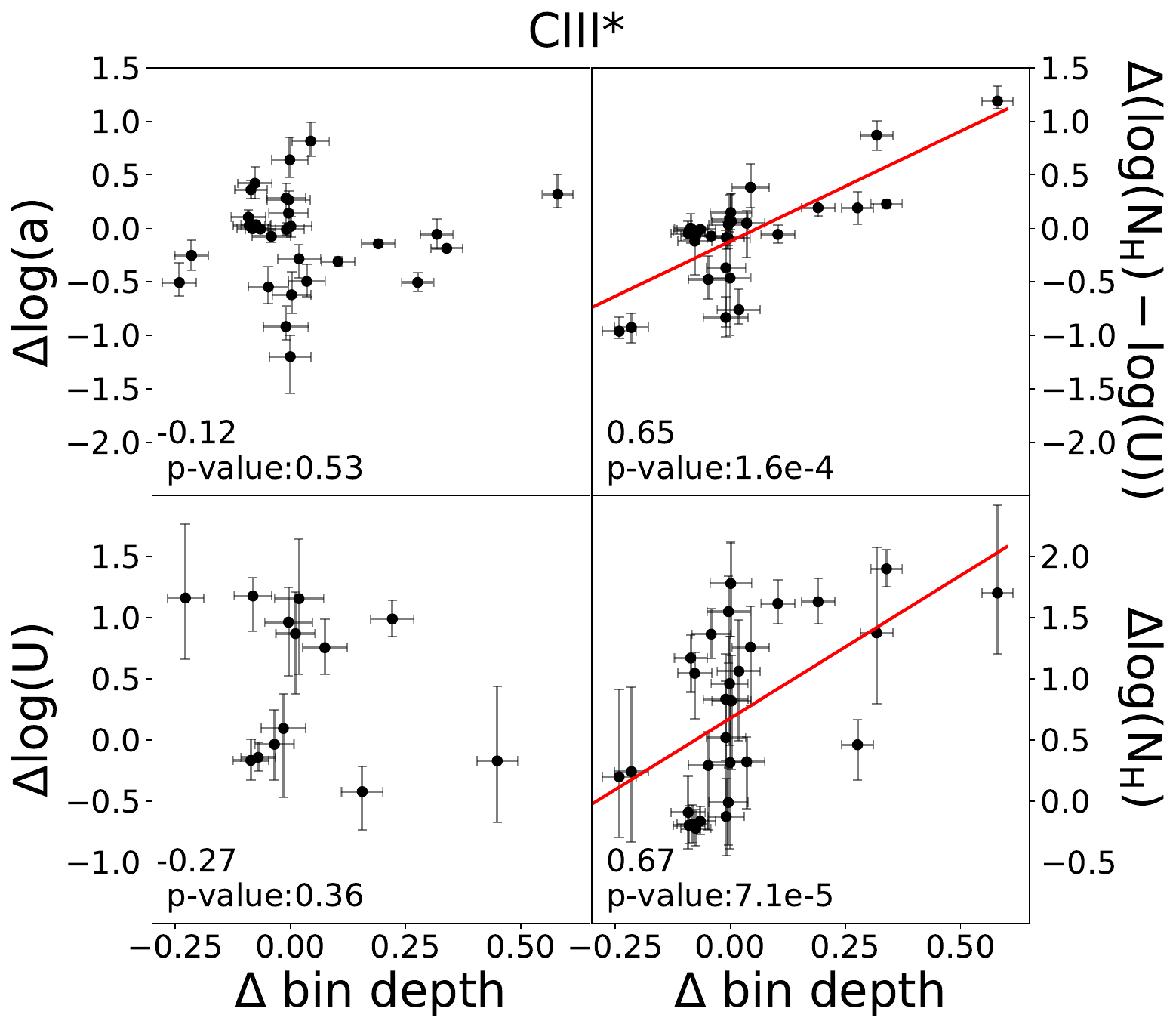}   \\
\includegraphics[scale=0.27,angle=0]{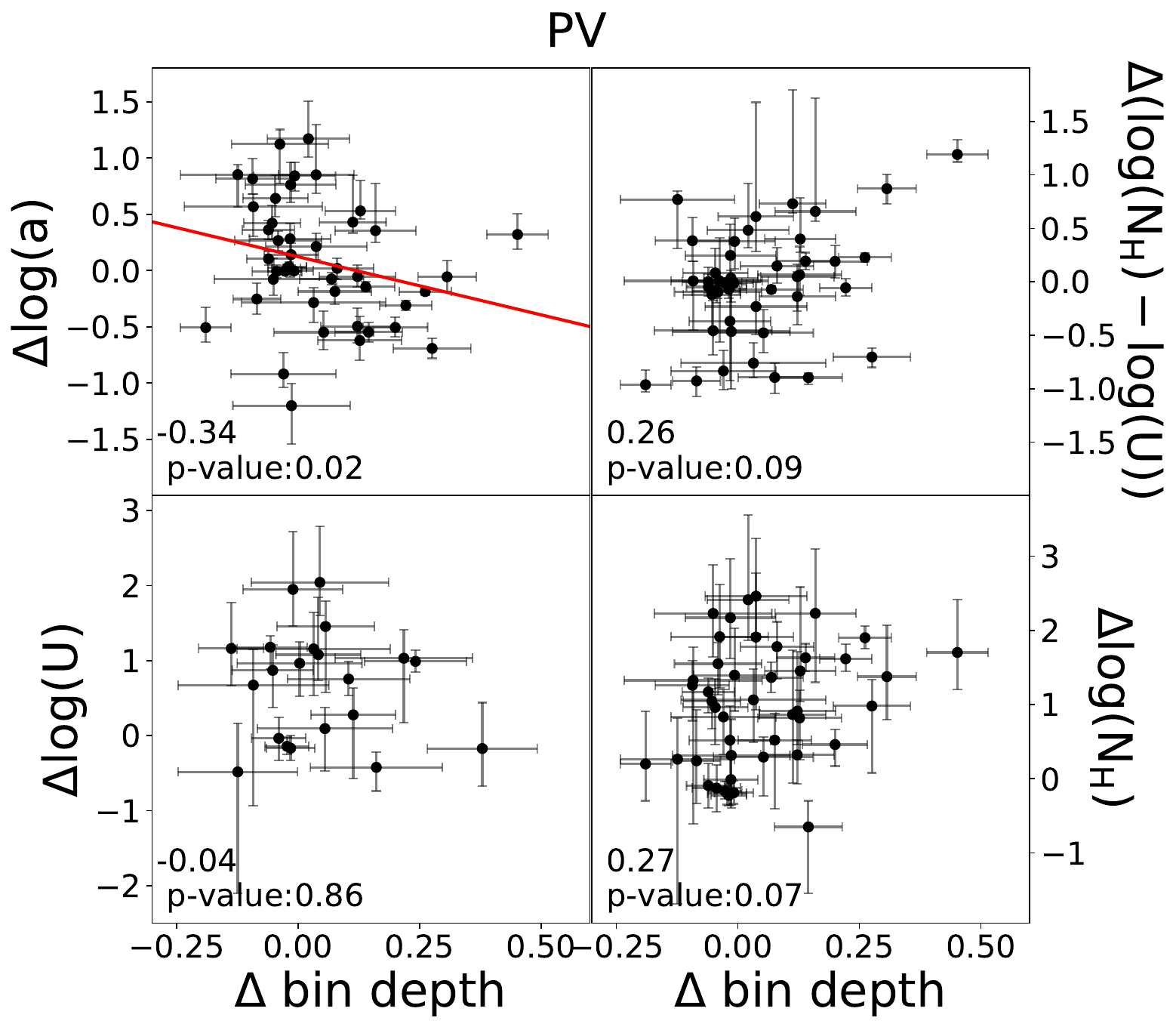}
\caption{Correlation plots between change in bin depth between epochs (x-axes) and change in physical parameter (y-axes) per ion. Spearman-rank correlation coefficients are provided with corresponding $p$-values. High ionization lines \CIV, \SiIV, \NV, and \PV\ all show a negative correlation with $\log(a)$, indicating a positive correlation with covering fraction. Low ionization line \CIII\ shows a positive correlation with column density and ${\log(N_{\rm H})-\log(U)}$. The best-fit lines are included in red for parameters showing statistically significant correlations for each ion. \label{fig:ion_emp}}
\end{center}
\end{figure*}

Using the final synthetic spectra from each single epoch fit, we produce bin-by-bin synthetic spectra for the \CIV, \CIII, \NV, \SiIV, and \PV\ lines and measure the change in bin depth per bin, {for each} ion between the same bins in different epochs. We examined the change in bin depth between each pair of epochs always with the form: later epoch - earlier epoch for each bin. Combinations of epochs are 2010-2013, 2010-2015, 2010-2017, 2013-2015, 2013-2017. We also calculate the change in each of the physical parameters (ionization parameter, covering parameter, $\log(N_{\rm H})-\log(U)$, and column density) per bin between pairs of epochs. There are fewer data points for ionization parameter ($\log(U)$) as all bins of a particular velocity group have the same $\log(U)$ value, and so we only include one data point per unique value of ionization parameter. The bin depth for those data points is the mean of the change in bin depth for all bins that share the same ionization parameter. We exclude 1 outlier data point with a change in $\log(U)$ of less than $-2$ as this data point came from the difference in ionization parameter from the high-velocity gas in 2010 and 2017, which may be poorly constrained due to a lack of absorption from \CIII\ and \PV\ in both epochs. We additionally exclude 3 data points with a $\log(U)$ change greater than 2.1, as all 3 data points corresponded to the changes in the lowest velocity component from the 2010 epoch where the 2010 data has almost no opacity. We use the {\tt scipy.stats} package to calculate the {Spearman}-rank correlation coefficient and corresponding $p$-value in each case. {The} results are shown in Figure~\ref{fig:ion_emp}. 

The high-ionization lines (\CIV, \SiIV, \NV, \PV) all show a negative correlation between the $\log(a)$ parameter and the change in bin depth (--0.88 for \CIV, --0.65 for \SiIV, --0.88 for \NV, and --0.34 for \PV) with very small $p$-values ($1.8\times10^{-15},\;1.3\times10^{-6},\;2.9\times10^{-10},\;0.02$ for \CIV, \SiIV, \NV, and \PV\ respectively) indicating that this correlation is strong and a decrease in covering fraction manifests as a decrease in line depth across all of the high-ionization lines. The low-ionization line (\CIII), which is unsaturated, did not show the correlation with $\log(a)$, but showed a positive correlation with column density (0.67 with a $p$-value of $7.1\times10^{-5}$) and $\mathrm{log(N_H)-log(U)}$ (0.65 with a $p$-value of $1.6\times10^{-4}$). These correlations may indicate that variability in different lines may be controlled by different parameters. {This analysis supports the result obtained from section~\ref{sec:multiepoch} that covering fraction is the primary driver of the variability observed.}

\subsubsection{Regression}\label{sec:regression}
The physical parameters fit with \simbal\ may not be independent, and so to further examine the relationship between the change in the depth of the absorption lines and the change in physical parameters, we fit a linear model to our results to look for a relationship between the BAL depth and physical parameters. 

A backward elimination method was used to determine which parameters are statistically significant { \citep[e.g.][]{2022bChoi}}. We {used an} iterative method to select features: first we fit a linear model to the data using the {\tt statsmodels.api} OLS function, then determined the maximum $p$-value for all parameters. If the maximum $p$-value was greater than 0.05 (and therefore not {considered }significant), {the parameter} was removed from the fit {and then this test was repeated until no parameters were left with $p$ values greater than 0.05}. We found that the only feature removed from \CIV, \CIII, and \PV\ was the ionization parameter, \logU, with \loga, \logNHU, and \logNH\ all being considered significant. For \SiIV, \logNH\ was also removed, leaving only \logNHU\ and \loga\ as significant. For \NV, \logNHU\ is removed in addition to \logU, leaving only \loga\ and \logNH\ as significant. For each ion line, once the significant features were selected, the dataset was then split into a training and test set with a test size of 20\% of the dataset. We fit a linear model to the training set, predict on the test data and calculate the coefficient of determination ($R_{\rm stat}^2$ value\footnote{Defined as one minus the ratio of the sum of residuals squared over the total sum of squares, $1-\frac{\Sigma^{n}_{i=1}(y_i-f(x_i))^2}{\Sigma^{n}_{i=1}(y_i-\bar{y})^2}$}), a statistic used to determine goodness-of-fit, using the test data. In each case we calculated an $R_{\rm stat}^2$ value greater than 0.9, indicating that for each ion our model is able to predict the change in BAL depth with reasonable accuracy. 

From this analysis, we confirm that \loga\ is the primary driver controlling the variability observed between epochs. In addition to \loga\ changing, either the column density or both the column density and the BAL thickness (as quantified by the \logNHU\ parameter) appear to be changing over time. We discuss the possible physical picture for this scenario in section~\ref{sec:discuss}.

\section{Physical Properties of the Outflow} \label{sec:other}

\subsection{Derived Outflow Properties}

The physical wind parameters that are directly fit by \simbal\ can be used 
to calculate derived physical properties of the outflow including total
 hydrogen column density, mass-loss rate, radius, and kinetic luminosity, 
 given in Table~\ref{table:derived_params}. We discuss in further detail 
 below how the properties were calculated and their implications.  
 For column density and radius, these values are given both by velocity 
 group and in total.    

\movetabledown=15mm
\begin{deluxetable*}{l l c c c c c}
\tabletypesize{\scriptsize}
\tablecaption{Derived parameters from the best-fit SimBAL model. \label{table:derived_params}}
\tablehead{\colhead{Parameter} & \colhead{Velocity Group} & \colhead{2010} & \colhead{June 2013} & \colhead{Dec. 2013} & \colhead{2015} & \colhead{2017}}
\startdata
$\mathrm{log(N_H)\;(cm^{-2})}$\tablenotemark{a}  & High & $22.31_{-0.29}^{+0.21}$ & - & - & - & $19.35_{-0.18}^{+0.32}$ \\
       & Medium  & $20.23_{-0.30}^{+0.26}$ & $21.33_{-0.19}^{+0.19}$ & $21.11_{-0.24}^{+0.23}$ & $21.41_{-0.42}^{+0.31}$ & $21.96_{-0.32}^{+0.18}$ \\
       & Low (1) & $21.34_{-0.48}^{+0.22}$ & $21.66_{-0.08}^{+0.09}$ & $21.47_{-0.11}^{+0.11}$ & $22.17_{-0.22}^{+0.13}$ & $22.39_{-0.14}^{+0.14}$ \\
       & Low (2) & $18.61_{-0.06}^{+0.06}$ & $21.36_{-0.07}^{+0.07}$ & $21.16_{-0.09}^{+0.09}$ & $21.73_{-0.49}^{+0.60}$ & $22.47_{-0.11}^{+0.12}$ \\
       & Total BAL & $22.76_{-0.20}^{+0.17}$ & $22.28_{-0.06}^{+0.06}$ & $22.11_{-0.08}^{+0.09}$ & $22.59_{-0.32}^{+0.51}$ & $23.24_{-0.12}^{+0.12}$ \\
       & Mini-BAL & $19.87_{-0.05}^{+0.08}$ & $20.13_{-0.05}^{+0.05}$ & $20.21_{-0.08}^{+0.08}$ & $20.12_{-0.09}^{+0.12}$ & $20.66_{-0.06}^{+0.07}$ \\
\hline
$\mathrm{log(R)\;(pc)}$ & High & N/A\tablenotemark{b} & - & - & - & $0.55_{-0.13}^{+0.10}$ \\
       & Medium  & N/A & $-1.58_{-0.15}^{+0.15}$ & $-1.60_{-0.24}^{+0.21}$ & $-1.12_{-0.28}^{+0.31}$ & $<-1.16$ \\
       & Low (1) & N/A & $-1.59_{-0.13}^{+0.12}$ & $-1.54_{-0.20}^{+0.17}$ & $-1.20_{-0.24}^{+0.25}$ & $-0.99_{-0.08}^{+0.07}$ \\
       & Low (2) & N/A & $-1.29_{-0.11}^{+0.11}$ & $-1.25_{-0.19}^{+0.15}$ & $-0.93_{-0.41}^{+0.38}$ & $-0.83_{-0.04}^{+0.04}$ \\
       & Mini-BAL & N/A & N/A & N/A & N/A & N/A \\
\hline
$\mathrm{log(\dot{M})\;(M_{\odot} \;yr^{-1})}$ & BAL & $-0.64_{-0.10}^{+0.14}$ & $-0.81_{-0.14}^{+0.15}$ & $-0.97_{-0.22}^{+0.20}$ & $0.10_{-0.30}^{+0.30}$ & $0.73_{-0.11}^{+0.11}$ \\
\hline
$\mathrm{log(L_{KE})\;(erg\;s^{-1})}$ & BAL & $42.92_{-0.11}^{+0.14}$ & $42.14_{-0.14}^{+0.15}$ & $41.97_{-0.21}^{+0.19}$ & $43.14_{-0.29}^{+0.26}$ & $43.80_{-0.11}^{+0.11}$ \\
\hline
$\mathrm{log(\dot{P})\;(dyne)}$ & BAL & $34.18_{-0.11}^{+0.14}$ & $33.69_{-0.14}^{+0.15}$ & $33.53_{-0.21}^{+0.19}$ & $34.64_{-0.29}^{+0.28}$ & $35.28_{-0.11}^{+0.11}$ \\
\enddata
\tablecomments{In each velocity section, values are the mean of the bins in each velocity group where the parameters were well sampled. Upper and lower limits represent the 95\% confidence interval sampled from the distributions of each parameter. Velocity ranges given are approximate for each epoch.}
\tablenotetext{a}{$\mathrm{log(N_H)+log(\frac{1}{1+a})}$}
\tablenotetext{b}{N/A is given for regions and parameters where the density could not be constrained due to a lack of \CIII\ absorption.}
\end{deluxetable*}

\subsubsection{Radius}
The radius of the BAL is calculated from the density and ionization parameter values from the \simbal\ fits{, and $Q$, the number of ionizing photons per second,} using the ionization parameter equation defined in section~\ref{sec:simbal}. The value of $Q$ is found by scaling a standard quasar SED (the soft quasar SED used to create the \simbal\ grids; \citealt{2011Hamann}) to best match the UV SED and photometry available for each epoch, {and }then integrating the photon flux for energies larger than 13.6 eV{. $Q$ is} estimated to be $\log(Q)=53.8-54.1${ (phot s$^{-1}$); the} range in values is due to the differences in the {normalization} for each epoch{ (see Figs.~\ref{fig:sed_uv}, \ref{fig:color_mag})}. 

{The radius of the medium and low-velocity gas has lower uncertainties because the density is better constrained by the presence of \CIII\ absorption.  To estimate the radius of the high velocity gas with no independent \CIII\ constraints, we assumed the same density values as the lower velocity gas for each epoch.  The lowest (mini-BAL) velocity absorbers have no independent density constraints, but are likely to have larger radius values than the BAL gas because the mini-BAL absorbing gas completely covers both the continuum and emission-line regions.}

\begin{figure*}[ht!]
\begin{center}
\includegraphics[width=16cm,scale=0.6,angle=0]{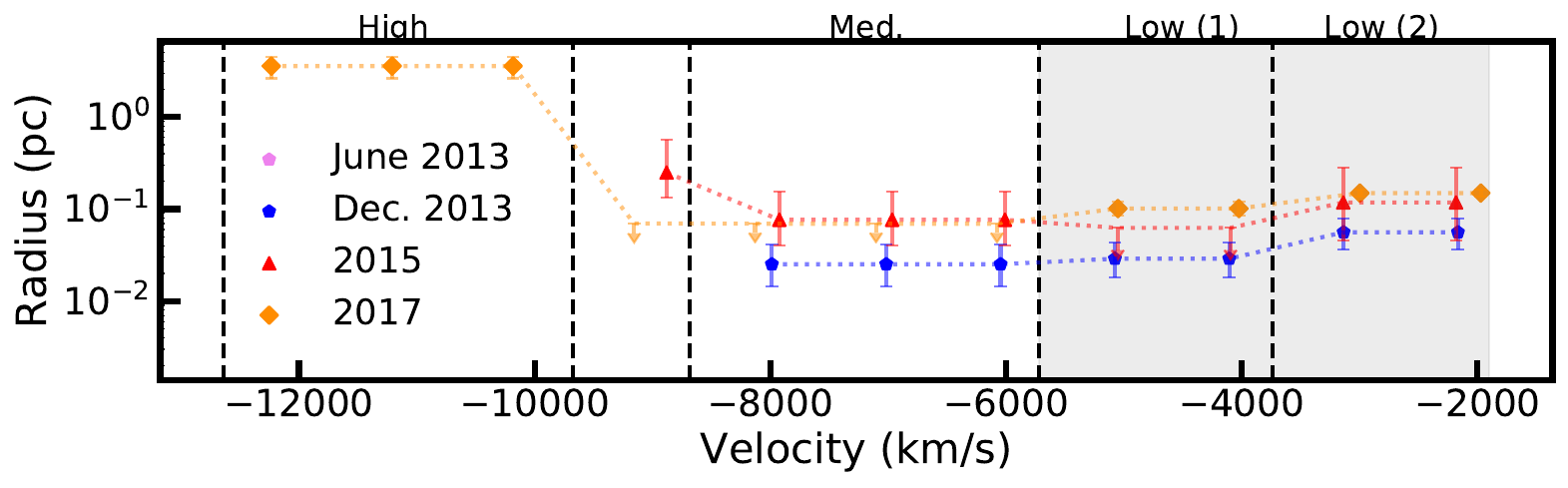}
\caption{Radius in pc for each velocity bin for the 2013 (filled pink and blue circles), 2015 (filled red triangles), and 2017 (filled orange diamonds) epochs. The radius in 2010 is not shown due to poor density constraints. Similarly, the mini-BAL radius is not shown due to poor constraints. However, the mini-BAL radius is likely to be larger because the mini-BAL fully covers the line emission and continuum. Points within the shaded region are points where the density is directly constrained by the \CIII\ absorption line. \label{fig:radiussingle}}
\end{center}
\end{figure*}

{The radius per velocity bin is presented in Figure~\ref{fig:radiussingle}. The density is measured per epoch and ionization parameter is measured per velocity group. The low and medium-velocity BAL gas has an average value across all epochs of 0.07 pc. The much larger radius calculated for the high-velocity component (3.5 pc) may be an artifact of the assumption that both components share the same density when they have much different values of ionization parameter.  If the high-velocity component has a higher density, the radius would be lower; a density value of $\rm log(n_H)\sim 8$ would put the high velocity gas at the same radius as the lower-velocity gas.  No radius value is presented for the BAL in the 2010 epoch or for the mini-BAL in all epochs due to the lack of \CIII\ absorption (the line is not in the bandpass in 2010) to directly constrain the density.  }

{For reference, t}he dust sublimation radius is estimated to be 0.01 pc calculated using the equation $R_{\rm sub}=0.2L_{46}^{1/2}$~pc from \citet{1993Laor} and a bolometric luminosity of $5.20\times10^{43}\mathrm{erg\;s^{-1}}$ \citep{Leighly2015}. With a radius estimate of $0.07\;\mathrm{pc}$ from \simbal\ models, this places the winds in the torus, consistent with \citet{Leighly2015} who found the radius to be in the {vicinity of the }torus with a distance of 0.17--1.47 pc based on {photometric} variability.

\subsubsection{Column Density}

We present the total hydrogen column density correcting for the partial covering of the BAL in Table~\ref{table:derived_params}. The log of the hydrogen column density varies between its lowest point in December 2013 (22.11 \cmsq) and its highest point in 2017 (23.24~\cmsq). Differences between epochs are minimal and there is no change within errors between the 2010, 2015, and 2017 epochs.

\subsubsection{Mass-Loss Rate}
Mass-loss rate is calculated per bin using $\dot{M}=8\pi\mu m_{\rm p} \Omega R N_{\rm H} v$ \citep{Dunn2009} in units of M$_{\odot}$~yr$^{-1}$ and the total mass-loss rate is the sum of the mass-loss rate per bin. The mean molecular weight, $\mu$, is assumed to be 1.4 amu and the global covering fraction, $\Omega$, is assumed to be 0.2 from the BAL population demographics \citep{2003Hewett}. Overall, the mass-loss rate (between $0.1$ and {5~M$_{\odot}$~yr$^{-1}$ across epochs) is high for this source given its low luminosity. In contrast, the mass outflow rate for the LoBAL SDSS J$085053.12+445122.5$ studied with \simbal\ was found to be 17--28~M$_{\odot}$~yr$^{-1}$ for a source that is $100\times$ more luminous \citep{Leighly2018}. However, a comparison to mass-loss rates found for AGN with comparable bolometric luminosities to WPVS~007 found mass-loss rates as high as $\mathrm{3.8\;M_{\odot}\;yr^{-1}}$ for the UV component of the outflow for source NGC~3516 and 4.61--8.25~M$_{\odot}$~yr$^{-1}$ for the UV outflow in NGC~3783 with bolometric luminosities of $\mathrm{1.4\times10^{44}\;erg\;s^{-1}}$ and $\mathrm{1.8\times10^{44}\;erg\;s^{-1}}$ respectively \citep{2012Crenshaw}. The sample studied in \citet{2012Crenshaw} studied UV and X-ray outflows from AGN; we compare our results to the listed total UV outflow component only, excluding the X-ray outflow. 

The mass accretion rate estimated using $\dot{M}_{\rm acc}=L_{\rm Bol}/\epsilon c^2$ assuming that WPVS 007 is radiating 10\% of its rest-mass energy ($\epsilon=0.1$) using a bolometric luminosity, $L_{\rm Bol}$, of $\mathrm{5.20\times10^{43}\;erg\;s^{-1}}$ \citep{Leighly2015} is $9\times 10^{-3}\;\mathrm{M_{\odot} \;yr^{-1}}$. With an Eddington accretion rate of $\mathrm{0.09\;M_{\odot}\;yr^{-1}}$, WPVS 007 is accreting at 10\% the Eddington limit. This would place the outflow rate between 20 and 543 times the accretion rate depending on the epoch of observation. The study of mass outflow rates of UV outflows for Seyfert 1 galaxies by \citet{2012Crenshaw} {found} mass outflow rates 10-1000 times the mass accretion rates for luminosities of 10$^{43}$--10$^{45}$~erg~s$^{-1}$ considering both the X-ray and UV outflows in their sample and 0.1--300 times the mass accretion rates considering the UV {outflow} components only (values for UV-only components were calculated{ using} available information in Tables 1 and 3 of \citet{2012Crenshaw}), indicating that the values found for WPVS 007 may not be unusual for a nearby AGN. We do note that we are comparing the UV BALs present in WPVS 007 to non-BAL UV {absorbers} in \citet{2012Crenshaw}, therefore the maximum velocity of the comparison AGNs are typically on the order of 1000 \kms\ or less, and some have black hole masses up to 10$\times$ higher than the mass of WPVS 007. Such a powerful outflow coming from a less massive black hole emphasizes how unique this system is.

\subsubsection{Kinetic Luminosity}
Kinetic luminosity {was} calculated per bin using $L_{\rm KE}=\frac{1}{2}\dot{M}v^2$ with the sum across all bins presented as the total kinetic luminosity in Table~\ref{table:derived_params}. WPVS 007 has a range in $\log(L_{\rm KE})$ of 41.97--43.80~erg~s$^{-1}$ (across epochs with 41.97~erg~s$^{-1}$ from the December 2013 epoch and 43.80~erg~s$^{-1}$  from the 2017 epoch) with corresponding $L_{\rm KE}/L_{\rm Bol}=0.017$--1.1. Comparing this value to {the kinetic luminosities of }NGC~3516 and NGC~3783 {from \citet{2012Crenshaw}} with luminosities roughly 3 times that of WPVS~007), we {found} that the outflow for WPVS~007 has a higher kinetic luminosity {than} NGC~3516 and NGC~3783 {which have} $\log(L_{\rm KE})$ values of 41.7 and 42.5 erg~s$^{-1}$, respectively {from their UV outflow components only (with much lower velocities than WPVS 007)}. The ratio of $L_{\rm KE}/L_{\rm Bol}$ for WPVS~007 is comparable to outflows from BALQs \citep[e.g.,][]{2013Borguet,2015Chamberlain,Choi2020}. This value is higher than the HiBAL source SDSS J$085053.12+445122.5$ studied with \simbal\ where the authors find $L_{\rm KE}/L_{\rm Bol}=0.0079$ (for the solar metallicity case;  \citealt{Leighly2018}). However, many more HiBAL objects need to be studied using \simbal\ in order to make a definitive comparison.

\subsection{Effective Covering Fraction}\label{sec:effcov}

As the power-law-partial-covering parameter is the most empirical and least intuitive of the \simbal\ physical parameters, we calculated the effective covering fraction per bin from the partial covering parameter. {The effective covering fraction is the fraction of the emission region covered. Since power-law partial covering must, by definition, assume that no region is truly uncovered, to calculate the total effective covering fraction, we assume some cutoff in optical depth, below which we assume there is no contribution to total covering. Effective covering fraction is taken to be the fraction of the surface area covered by gas with an opacity greater than 0.5. The $\tau>0.5$ criteria is given in \citet{2005Arav} as a good cutoff to estimate the effective covering fraction. Optical depth as a function of velocity in the power-law partial covering case is defined as $\tau=\tau_{max}x^{a}$ where $\tau_{\rm max}=2.654\times10^{-15}f\lambda N_{ion}(v)$ \citep{1991Savage} and $f$ is the oscillator strength of the transition, $\lambda$ is the wavelength (\AA), and $N_{ion}(v)$ is the ion column density ($\mathrm{cm^{-2} (km\;s^{-1})^{-1}}$). This calculation was done for the high ionization lines of \CIV, \SiIV, \NV, and \PV, as well as the low-ionization lines \CII\ and \CIII\ to examine changes in effective covering fraction as a function of velocity, epoch, and ion line as was done in Figure 16 of \citet{Leighly2019}.} 

In Figure~\ref{fig:effcovsingle}, we show the effective covering fraction {per velocity bin} as a function of transition and time from 2010 to 2017. These results show that the high ionization lines cover more of the continuum than the low ionization lines. The covering fraction was higher for the medium velocity gas in 2010 and 2017 when the BALs were deepest and the high-velocity component was present. The effective covering fraction was lowest for the medium and low velocity gas in 2015 when the high-velocity BAL component was absent and the BALs were the shallowest. At the lowest velocities, the effective covering fraction does not change significantly from 2013 to 2017 and even appears to increase in 2015, although these bins overlap with emission lines so small changes between epochs could reflect changes in the emission lines and not a change in absorption. {For the bins around 8,000 \kms\ there is a change in covering fraction of \SiIV\ from $<0.1$ in 2015 to 0.4 in 2017, whereas we see almost no change in \SiIV\ for the low-velocity bins between the two epochs. Unlike previous work with WPVS 007, we find absorption at low velocities in 2015 due to our finding that the line emission is uncovered.} The lowest velocity gas seemed to disappear from \CIV\ and \PV\ in 2010, but the \SiIV\ component remained well-covered. {Overall, we observe small changes in effective covering align with small changes in BAL depth for a particular velocity bin and large changes in effective covering align with regions of the BAL where larger changes were observed.}

\begin{figure*}[ht!]
\begin{center}
\includegraphics[scale=0.35,angle=0]{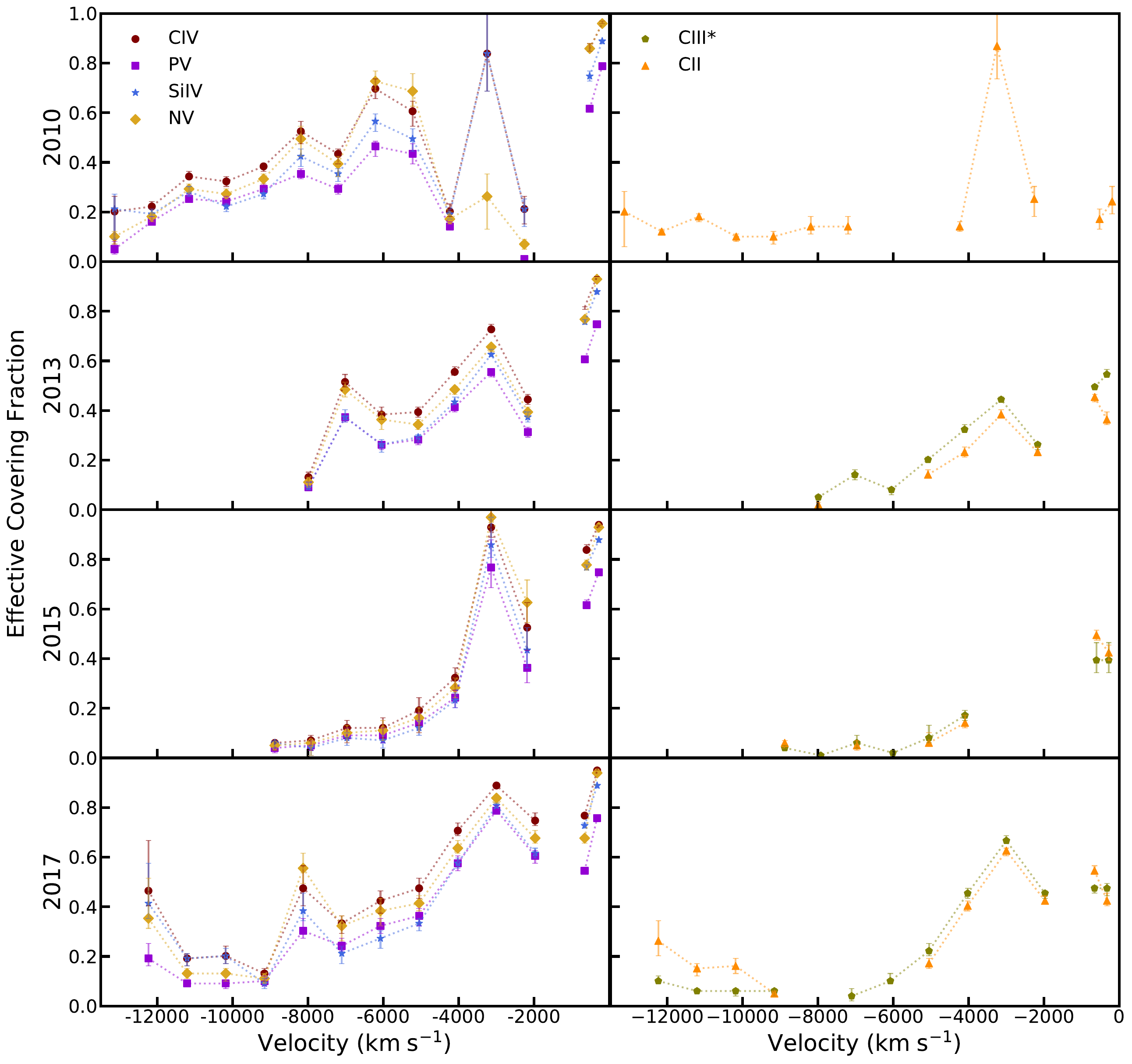}
\caption{Effective covering fraction as a function of time, velocity, and ion. In general we see high-ionization lines (left) have a higher covering fraction than low-ionization lines (right) and that the low-velocity gas has a higher covering fraction than the high-velocity gas {for all ions}. The two points at the lowest velocities in each plot represent the mini-BAL covering fraction, which is consistent with having a covering fraction equal to or greater than the lowest velocity BAL gas. \label{fig:effcovsingle}}
\end{center}
\end{figure*}
 
\subsection{Mini-BAL Outflow}

The mini-BAL parameters are the same within errors between epochs for \logU, \logNHU, \logNH, and \loga, consistent with the observed lack of variation in mini-BAL depth between epochs. This gas shows lower ionization parameter (\logU average of $-0.79$ across all epochs) compared to the BAL gas (\logU\ average of 1.27 across all epochs and velocities), lower column density (average \logNH\ of 20.6~\cmsq) compared to the total column density of the BAL (average \logNH\ of 22.6~\cmsq), and higher covering fraction (average \loga\ of 0.71 for the mini-BAL) than the BAL (average \loga\ of 1.5 across all epochs and velocities). We determined that the BAL does not cover the line-emitting region, whereas the mini-BAL fully covers the emission-line region. The difference in covering of the line and continuum from the BAL gas, together with the large covering fraction, suggests that the mini-BAL is an outflow separate from the BAL gas, existing on larger scales and likely farther away (although this cannot be confirmed {without the density constraint needed to calculate $R$}). An estimate for the ratio of the radius of the mini-BAL outflow to the radius of the BAL outflow for different epochs{, }assuming that the mini-BAL has the same density of the BAL for a particular epoch{,} shows that the mini-BAL could be around 10 times the distance of the BAL outflow. 

\section{Discussion} \label{sec:discuss}

With \mbh$=4.1\times10^{6}$ \msun\ and a small radius for the BAL outflow, $R_{\rm BAL}\sim0.07$ pc (this work), WPVS 007 has a significantly smaller mass and outflow size than the BALQs at $z\sim2$ that are the principal target of variability studies to date{ \citep{2009Leighly}}.  Previous work has estimated the lifespan of a typical BAL to be decades or longer \citep{Gibson2008,2012Filiz,FilizAk2013}.   However, the low redshift ($z=0.028$, with correspondingly little time dilation) and smaller size means that variability studies of WPVS 007 on timescales of years have the ability to probe dramatic BAL changes that might take decades to see with larger, more massive, and higher redshift samples.  Furthermore, variability studies of typical BALQs (with masses of $\sim10^9$ \msun) have shown that the amount of variability increases with increasing time between observations \citep[e.g.,][]{Capellupo2013}. As seen in Figure 1 of \citet{Capellupo2013}, the fractional change in maximum absorption-line depth of 24 \CIV\ BALs on timescales up to 10 years is no more than 0.5. We observe comparable fractional changes in absorption-line depth of 0.4 for WPVS 007 within only one year. 

\subsection{Comparison of Outflow Properties with Previous Work} \label{sec:discuss_outflow}
With our new multi-epoch \simbal\ analysis with { simultaneous fits of the continuum and several absorption lines}, we are in a position to better constrain the physical properties of the outflow, and so we compare this work to previous studies of WPVS 007.
From the \simbal\ fit results, we found that the ionization parameter of WPVS 007 is high, up to \logU=1.88 (as seen in 2015).  This ionization parameter is significantly higher than previous constraints on the ionization parameter from studies of this object. For example, \citet{Li2019} gave a maximum value for $\log(U)$ of 0.0.  

The \simbal\ analysis gives us confidence in our results, as we simultaneously fit multiple lines as a function of velocity.  The strong, high ionization lines of \PV\ and \NV\ in particular are powerful diagnostics.  
The model fits presented here show strong \NV\ and \PV\ BALs, with a \PV\ {apparent }optical depth almost matching that of \CIV\ at the lowest velocities. These lines together imply a large ionization parameter and a large column density \citep[e.g.,][]{1998Hamann} because of the low cosmic abundance of \PV\ ($\mathrm{P/C=0.00093}$; \citealt{2007Grevesse}). 
Specifically, we find that the ionization parameter varies from \logU=--0.28 to \logU=1.88 for bins that cover the \PV\ BAL.  Our value is consistent with the previous analysis of the 2003 FUSE observations by \citet{2009Leighly} that gave the lower limit of  \logU$\gtrsim 0.0$ from the strong, almost saturated \PV\ absorption line and saturated \NV\ line.  In addition, the assumption of \citet{Li2019} that the \NV\ BAL is weak is not consistent with our spectral-fitting results. Assuming complete covering {of the continuum and emission lines } leads to a significant underestimate of the amount of absorption from the \NV\ line. BALs not fully covering emission lines have been observed before \citep[e.g.,][]{1999Arav,2012Borguet,2022Choi}. 
With a robust determination of the ionization parameter and the density constraints from the \CIII\ line, we used the definition of ionization parameter to calculate the radius of the outflow over the velocity bins where the \CIII\ feature is present.  The average radius across velocity regions (assuming the density is uniform) and epochs (excluding the high-velocity gas in 2017 and the 2010 epoch) is 0.07 pc. The initial launching radius of the outflow can be estimated from ionization and kinematic arguments (based on assuming the Keplerian velocity at the launch radius and terminal velocity of the outflow are comparable), and gives much smaller values of $\sim10^{-4}$~pc \citep{2009Leighly,Li2019}.  Alternately, \citet{Leighly2015} give an estimate for the outflow radius of 0.17--1.47~pc based on the variability, with a {torus radius of $R_{\tau K}\sim 0.036$ pc.  Our value for the outflow is thus also } consistent with being found in the vicinity of the torus.
 
\subsection{What Causes the BAL Variability in WPVS 007?}
Understanding the physical reason behind BAL variability offers promise for understanding more about the geometry and structure in quasar outflows.  With the use of \simbal, we have the unique opportunity to study the difference between the physical properties of the gas over time. To date, many studies of variability use an observed change in equivalent width of the \CIV\ line (or \CIV\ and \SiIV) to inform the understanding of the cause of the variability in the BAL \citep[e.g.,][]{Capellupo2011,Capellupo2012,Capellupo2013,2012Filiz,FilizAk2013,2014FilizAk}.  However, studies such as \citet{Lundgren2007} have emphasized the importance of having lines other than \CIV\ {(which is often saturated)} when studying variability.

In the case of WPVS 007, we find this to be true; had we studied only the \CIV\ line, we would not have been able to distinguish between a change in ionization parameter, column density, or covering fraction. In the multi-epoch fits in section~\ref{sec:multiepoch}, all tested scenarios were able to fit the \CIV\ line for the December 2013 and 2017 epochs (see Figure~\ref{fig:multi_epoch_acxcomp}).  {As mentioned above, the presence of \PV\ in the spectrum, given the low abundance of phosphorus compared to carbon and its high ionization potential,} is an indicator of high column density and high ionization parameter \citep[e.g.,][]{2009Leighly,Leighly2018,2014Capellupo}.  In this parameter regime, the \CIV\ absorption line will always be highly saturated \citep{2012Borguet}, and therefore large changes in some physical properties of the outflow may show little effect in the \CIV\ EW.  In particular, saturated lines -- as \CIV\ is whenever a \PV\ BAL is present -- are insensitive to changes in ionization parameter \citep{2017McGraw,2019Vivek}.  For WPVS 007, it was only with the combined simultaneous fit of several lines, including \PV, \NV, \SiIV, and \CIV, that a change in covering fraction was revealed to be the driving source of variability between epochs.  From Figure~\ref{fig:multi_epoch_acxcomp}, the models where the ionization parameter or column density were allowed to vary could not produce the full depth of the \PV\ BAL. \CIII\ also was critical in the analysis as the only density-sensitive line. Like \PV, \CIII\ is not saturated, but is also not present at all velocities where \CIV\ is observed. 

Previous studies of BAL variability have linked the cause of variability to one of two causes: (1) a change in ionization of the gas caused by a change in strength or shape of the ionizing continuum or (2) a change in absorption caused by gas moving transverse across our line of sight with the quasar, the ``cloud-crossing'' scenario. The origin of the cause of variability is frequently used to determine the physical conditions of the gas. If a change in ionization is causing the observed variability, then the timescale between observations can be used to estimate the electron density and then the radius \citep[e.g.,][]{1993Barlow}. If variability is caused by cloud crossing, the timescale and degree of change in the depth of the BAL can be used to estimate the crossing speed and therefore the radius \citep[assuming Keplerian motions, e.g.,][]{Capellupo2013}. Using \simbal, the constraints on ionization parameter and density that allow us to calculate the radius are obtained directly from the MCMC fits.

Overall, it is unlikely that a change in continuum flux is driving the observed changes in the BALs. A changing ionization parameter causing the observed changes in absorption would indicate that the variability is {likely} being caused by a changing continuum.  We found no support for a change in ionization parameter driving the variability in the BALs as discussed above through our use of simultaneous fitting as well as poor correlation between the change in ionization parameter and change in depth of the BALs.  {This interpretation of the variability differs from that proposed by \citet{Li2019}. \citet{Li2019} examined patterns between the WISE infrared photometry, the \swift\ photometry, and the equivalent widths of the \CIV\ and \SiIV\ lines to find that the cause of the variability is due to a change in the ionizing continuum instead of a change in a line-of-sight absorber. They were able to reproduce the observed Swift light curve using stochastic changes to the emission from the disk. They further calculate the concordance index (absorption line EWs are assigned a $+1$ if the EW and continuum change in the same direction and $-1$ for the opposite case) to compare the stochastic changes in the disk to the variation in the UV \SiIV\ and \CIV\ BALs. In comparing the UV luminosity and BAL EW for the same 5 epochs of spectroscopy used in this analysis, \citet{Li2019} found a positive concordance index (the luminosity and BAL EW vary in the same direction) in all cases for \CIV\ and most cases for \SiIV. This method has been used to link a change in continuum with a change in absorber optical depth before \citep[e.g.,][]{2015Wang}. We observe the same stochastic changes in the photometry (see Figs.~\ref{fig:photo} and \ref{fig:color_mag}) due to variation in the accretion disk, but disagree that these stochastic changes are the reason for the observed variation in BAL EWs. Using \simbal, we are able to test these underlying assumptions that a changing EW is controlled by a change in the ionizing continuum. Given the saturation of the \CIV, the EW of this line is not sensitive to changes in ionization state. Moreover, the high ionization parameter required by the presence of \PV\ absorption means that the concordance scenario proposed by \citet{Li2019} predicts a decrease of \CIV\ EW with an increase in luminosity contrary to what has been observed.}

{ Previous studies have linked coordinated variability across all velocities of a BAL to a change in ionization state, whereas a change in a portion of the BAL is often linked to a change in covering fraction \citep[e.g.,][]{Capellupo2013}. Although coordinated variability fits more naturally to a change in ionization parameter (a change in continuum would globally affect the outflow), this is not the case for saturated BALs, where covering fraction controls the apparent opacity of saturated lines \citep[see][]{2011Hall,2014Capellupo}. Much like the case of Q1413+1143 observed by \citet{2014Capellupo}, WPVS 007 shows a saturated \CIV\ line with variability in \CIV, \SiIV, and \PV, which indicates a change in ionization state of the gas cannot cause the variability, even if there are apparently coordinated changes across the BALs, because covering fraction is controlling the depth. Furthermore,  though variations across the BALs appear to vary together, the changes across the BALs for WPVS 007 are not as coordinated as a function of velocity as they initially appear. Although the overall depths of the lines change, there are smaller changes along the BAL. As a function of velocity, not all components of the \CIV\ line vary uniformly (e.g., the depth at the blue edge of the \CIV\ line is changing separately from the red edge). For example, in 2015, most of the BALs seem to have gotten weaker in tandem. However, SimBAL analysis (which considered separately the high and low velocity BAL components) identified more absorption at low velocities in 2015 than previously modeled.  Once the emission-line covering fraction was accounted for (with the low velocity BAL gas not covering the emission lines), the low-velocity gas component did not disappear in 2015 as previously thought.}  

From fitting the continuum emission, we found that there is no significant change to either the power-law component or emission-line parameters between 2013 and 2017, but there is a significant change to the BAL absorption between these epochs. Furthermore, the continuum is different between 2015 and 2017, with the 2015 continuum being significantly redder, but the ionization parameter of the BAL gas between 2015 and 2017 is consistent. Although the \swift\ photometry shows variation epoch-to-epoch, these photons are longer in wavelength than the continuum photons that would be ionizing the gas in the outflow. We measured the continuum flux levels during the 2013 and 2017 HST observations by sampling the median flux in a line-free region of the continuum (1450~\AA). We observe no change within errors between 2013 and 2017, while at the same time we observe a dramatic change in the BAL lines. The 2010 and 2015 epochs show the strongest and weakest continuum flux levels, respectively, consistent with the {\em Swift} photometry (see Fig.~\ref{fig:color_mag}). This indicates that there are changes in the continuum over time, but those changes are not driving the BAL variability between all epochs. For WPVS 007, we can therefore reject scenario (1), a cause in ionization parameter, and now consider the cloud-crossing scenario. 

In previous variability studies, a change in covering fraction has been synonymous with the cloud-crossing scenario, where an absorber passes across the continuum-emitting region along the line of sight \citep{Gibson2010,2008Hamann,2012Filiz,2011Hall,Capellupo2013,2015McGraw,2017McGraw,2012Vivek,2016Vivek}. {In particular, \citet{2014Capellupo} found a case of a \PV\ BAL showing variability in the \PV, \CIV, and \SiIV\ lines and determined the cause of the variability to be cloud-crossing.} A change in absorption-line structure due to cloud crossing would produce a change in partial covering parameter, \loga, and possibly also a change in column density in \simbal. Although we have ruled out a change in column density as the main driver of the variability; the results of the regression analysis discussed in \S\ref{sec:regression} imply that column density may also be a contributing factor determining the change in depth of the BALs. The cloud-crossing scenario for the case of WPVS 007 is shown in the middle panel of Figure~\ref{fig:diagram}.

\begin{figure*}[ht!]
\begin{center}
\includegraphics[width=16cm,scale=0.6,angle=0]{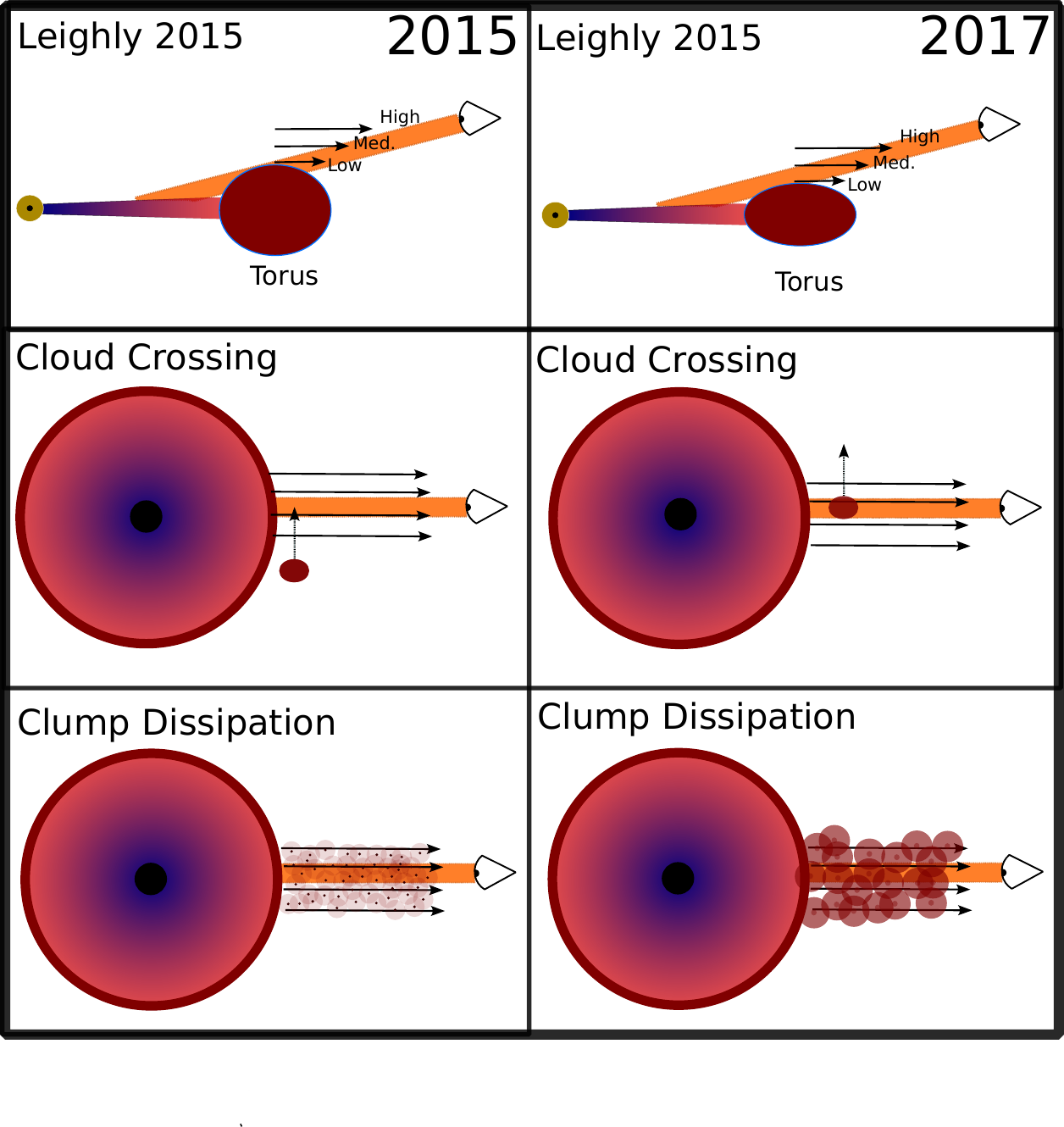}
\caption{Different scenarios for the changing outflow of WPVS 007. In the top panel, we show a cross-section of the scenario described in \citet{Leighly2015} where the scale height of the torus changes with time and our line of sight with the continuum is occasionally eclipsed by the torus (left, 2015). The outflow is launched from the top of the torus and in epochs where the scale height is low the high-velocity gas has a higher covering fraction because the line of sight passes more directly through the high-velocity gas. In the middle panels, we show aerial views of the cloud-crossing scenario where a cloud of material passes along the line of sight. The bottom panels show an aerial view of the clump dissipation scenario where a change in partial covering is caused by a change in the distribution of material along the line of sight. In 2015 (left) where the covering fraction is lower and the optical depth of key lines is lower, we see dense knots where the opacity is high surrounded by more diffuse gas with lower opacity. In 2017 (right), there is a more even opacity across the individual ``clouds'', resulting in a higher covering fraction and observed larger optical depth of key lines. The wind in the bottom panel is made to resemble Figure 12 of \citet{Leighly2019}. \label{fig:diagram}}
\end{center}
\end{figure*}

We can calculate the expected time for the BAL to completely cross the continuum, assuming Keplerian motion of the gas around the black hole. Using a radius of 0.07~pc, the speed of the gas is about $\mathrm{500\;km\;s^{-1}}$, from ${v_{\rm cross}=\sqrt{GM/R}}$ where $G$ is the gravitational constant, $M_{\rm BH}$ is the mass of the black hole and $R_{\rm BAL}$ is the outflow radius. Given the size of the accretion disk at 1550~\AA\ is estimated to be $R_{\rm 1550}=(3.4-7.9)\times10^{-4}$~pc \citep{Leighly2015}, it would take between $\sim10$ months and $\sim1.6$ years to cross the continuum at $R$. This crossing time is shorter than the time between the luminosity peak in 2010 and the occultation event in 2015.  We can use the cloud-crossing method outlined in \citet{Capellupo2013} to calculate the crossing speed from the change in depth of the BAL, assuming the absorber and continuum are uniform and circular. Considering all bins and epochs, the crossing speed would range from 6.6--423 km~s$^{-1}$, which leads to an outflow radius of 0.1--400 pc. At the low end, this is only marginally higher than the 0.07 pc radius determined using \simbal, making the cloud-crossing scenario a possible cause of the variability observed in WPVS 007. 
 
\citet{Leighly2015} speculate that the outflowing gas may be ablated material driving from the torus the same way ``wind ablates spray from the crest of a wave".  The outflow {would therefore be }the upper edge of the torus; this is consistent with the constraints on $R_{\rm BAL}$ from \simbal\ and $R_{\rm torus}$ from \citet{Leighly2015}.  It is widely accepted that the torus must be clumpy \citep[e.g.,][]{2015Netzer}.  Given this picture, \citet{Leighly2015} proposed that the variability is caused by the changing scale-height of the torus that on occasion more completely eclipses our line-of-sight. The changing torus scale-height model assumes that our line-of-sight skims the edge of the torus, and as the torus rotates, our view of the quasar is either relatively unobscured when the torus has a low scale-height (in the case of 2010 where the BAL was strong, the photometry blue, and the emission lines broad), or highly obscured when the torus has a high scale-height (in the case of 2015 where the BAL was weak, the photometry red, and the emission lines narrow). The outflow is clumpy and embedded in a larger region of gas and dust that controls the variability on longer timescales. A changing covering fraction would be consistent with the \citet{Leighly2015} picture. The changing torus scenario is shown in Figure~\ref{fig:diagram} and in Figure 5 of \citet{Leighly2015}. 

Alternatively, if the outflow is located beyond the location of the torus, when the scale-height of the torus is low (2010, 2013, 2017), it does not block{ any of} the continuum, and {therefore the} angular size of the continuum is larger{ from the point of view of the outflow.  (Note that we are assuming that the torus is completely opaque in this picture.)}  In contrast, a high torus scale height (2015) blocks part of the continuum.  In this case, the changing covering fraction results from the changing angular size of the continuum, and not a cloud crossing the observer's line of sight. {When t}he angular size of the continuum as viewed from the outflow is larger, {more of the outflow can cover the continuum, and so }the covering fraction is larger; the changing scale-height model is the geometry that allows this to happen. 

Alternatively, a change in covering fraction without a change in column density could be consistent with the formation and dissipation of clumps of material along the line of sight. Due to the way that \simbal\ uses power-law partial covering, a change in power-law partial covering could be interpreted as a change in how diffuse clumps are along our line of sight (see Figs. 12 and 20 in \citealt{Leighly2019}). We have used this idea from \citet{Leighly2019} to construct a diagram of what could be {occurring} in WPVS 007 (see the bottom panel of Figure~\ref{fig:diagram}). In this picture, the smaller, dense portions of clumps will have high opacity and {lower} covering fraction. However, the steep opacity profile would mean low opacity for other (larger) regions within the wind; we propose that this could be the case for the wind in 2015 when the covering fraction of the high and medium velocity gas is low. With a shallow opacity profile across clouds, we would see increased optical depth and therefore a higher covering fraction across all velocities (this would be the case for 2017 as we demonstrate in Fig.~\ref{fig:diagram}). 

We calculated the dissipation timescale of a clump of material, assuming that the high velocity, medium velocity, and low-velocity gas all represent distinct single ``clouds" within the outflow. The diameters of the clouds are estimated using $R\sim N_H/n_H$ and the dissipation timescale is assumed to be $t\sim R/v$, where $v$ is the maximum velocity of the gas \citep[e.g.,][]{2013Hamann}. We found that the dissipation timescale has a large variation between all epochs and velocity groups. The shortest dissipation timescales occur for the high-velocity gas in 2017, where dissipation occurs on the order of 3 hours. The medium velocity gas across all epochs shows dissipation timescales on the order of hours for the 2013 observations, which is inconsistent with the lack of observed variability between June and December observations. The longest timescales we find are for the low-velocity gas in 2017 with dissipation timescales of almost 2 years. Considering that we see little variation for the low-velocity gas from 2013 to 2017, this is also inconsistent. The assumption that entire portions of the outflow consist of single large clumps of gas is very limiting and likely not physical. \simbal\ itself uses a partial covering approach that assumes that the wind is made up of an ensemble of small clouds. However, the assumption that the clouds are small would further reduce the dissipation time of the clumps. MHD simulations of disk winds have shown evidence for the formation and dissipation of high-density knots that form from instabilities in the wind every $\sim$3 years \citep{2000Proga}, however these simulations were for {line-driven accretion} disk winds in a much larger $10^8$~\msun\ system. The outflow of WPVS 007 is highly unusual in its dramatic variability and may make an ideal candidate for {evaluating} future models of AGN outflows.

\section{Summary and Future Work} \label{sec:summary}
{ WPVS 007 is a low-redshift (z=0.028) NLS1 with a small black hole mass ($\mathrm{10^6\;M_{\odot}}$) that shows high-velocity broad-absorption lines characteristic of BALQs. Due to its small mass and more compact size, the outflow in WPVS 007 varies on a faster timescale than its much more massive quasar counterparts making it an excellent lab for studying BAL variability.}
We present the analysis of the BAL variability of NLS1 galaxy WPVS 007 using forward-modelling and spectral synthesis code \simbal. By fitting five epochs of HST COS observations taken between the years of 2010 and 2017, we determine that a change in covering fraction of the continuum is controlling the variability observed in the BALs. Specifically, we found a correlation between the change in partial covering parameter (used in place of covering fraction) from \simbal\ analysis and the change in depth of individual BAL lines. The partial covering parameter was determined to be the main driver of the variability across high-ionization lines including \CIV, \SiIV, \NV, and \PV. This result was confirmed using pairwise fits of epochs from December 2013 and 2017 using \simbal, where varying the partial covering parameter alone was able to fit the variability in all 3 epochs, December 2013, 2015, and 2017. A change in ionization parameter or column density alone could not fit the lines across these epochs. 

{This analysis was the first \simbal\ variability analysis performed, and the first time \simbal\ was used to fit more than one absorption spectrum simultaneously. 
\simbal\ is therefore a powerful tool for linking a physical change in the gas to observed variation within the BALs and paves the way for future variability studies of BALQs that show different characteristic variability that may arise from other changes in physical conditions.}

Although our solution is well constrained, we are limited by a lack of density-constraining absorption lines. The high-velocity gas did not show absorption for key diagnostic lines such as \CIII, limiting our ability to draw conclusions about the nature of the high-velocity gas component. We used a single density parameter constrained by gas at low velocities to describe all gas in the system, when the high-velocity gas may have different physical conditions. Multiple epochs of observation of density-sensitive lines like \SIV\ (observed in 2003 with FUSE but not {present in the band pass }in any of the HST COS epochs) would help constrain the density in general, but may not help in modelling high-velocity components that only show absorption in \CIV. 

Only one observation was taken with COS during its peak in the UV light curve, in 2010. The 2010 observation happened to be taken in a mode that did not observe critical BALs necessary to perform a proper fit, including the \CIII\ BAL, \Lyalpha, and \NV\ BAL in full. There continues to be regular monitoring of the UV photometry with \swift, and so an ideal observation to better understand the nature of the outflow would be to observe WPVS 007 in the high state with HST COS in a mode to cover all absorption lines between 1065 and 1600\AA\ along with simultaneous deep X-ray observations to constrain the {shape of the ionizing continuum and the absorber total } hydrogen column density.

\begin{acknowledgements}

We thank Sam Barber and the anonymous referee for constructive feedback that improved the paper. This work was made possible by the facilities of the Shared Hierarchical Academic Research Computing Network (SHARCNET) and Compute/Calcul Canada. SG and KG acknowledge the support of the Natural Sciences and Engineering Research Council of Canada (NSERC). Nous remercions le Conseil de recherches en sciences naturelles et en g\'enie du Canada (CRSNG) de son soutien. KG acknowledges the support of the Ontario Graduate Scholarship Program. Support for \simbal\ development and analysis is provided by NSF Astronomy and Astrophysics grants No. 1518382, 2006771, and 2007023. Some of the computing for this project was performed at the OU Supercomputing Center for Education \& Research (OSCER) at the University of Oklahoma (OU).  This research is based on observations made with the NASA/ESA Hubble Space Telescope obtained from the Space Telescope Science Institute, which is operated by the Association of Universities for Research in Astronomy, Inc., under NASA contract NAS 5–26555.  Based on observations with the NASA/ESA Hubble Space Telescope obtained Space Telescope Science Institute, which is operated by the Association of Universities for Research in Astronomy, Incorporated, under NASA contract NAS5-26555. Support for Program number HST-GO-14693.001-A was provided through a grant from the STScI under NASA contract NAS5-26555. KML \& HC acknowledge support from HST-AR-15035.001-A.
We would like to thank the PI Brad Cenko for his continuous support of observing WPVS 007 and approving our observing requests, and the \swift\ Science Operations team for executing the observations. This research has made use of the XRT Data Analysis Software (XRTDAS) developed under the responsibility of the ASI Science Data Center (ASDC), Italy. This research has made use of data obtained through the High Energy Astrophysics Science Archive Research Center Online Service, provided by the NASA/Goddard Space Flight Center.

\end{acknowledgements}
 
\bibliography{wpvs007_kaylie_astroph}

\end{document}